# Rethinking domain generalization in medical image segmentation: One image as one domain

Jin Hong[1, *], Bo Liu[2]

1. School of Artificial Intelligence, Nanchang University, Nanchang, 330031, China

2. School of Mathematics and Computer Science, Nanchang University, Nanchang, 330031, China

E-mail: hongjin@ncu.edu.cn; liuboncu@email.ncu.edu.cn

* Correspondence should be addressed to Jin Hong

**Abstract**: Medical image segmentation faces persistent challenges from domain discrepancies, particularly when integrating data across institutions. Notably, variations within single centers—driven by differences in scanner configurations or imaging protocols—often create shifts as substantial as those observed between facilities. To overcome these limitations, our work introduces the One Image as One Domain (OIOD) hypothesis, treating each image as an independent domain to enable adaptable generalization. Building on this principle, we present a unified disentanglement-based domain generalization (UniDDG) framework that eliminates dependency on domain labels while remaining adaptable to both single- and multi-source scenarios. The architecture maintains consistent complexity regardless of the number of source domains and reduces to the content encoder and segmentation head at inference. UniDDG operates by isolating anatomical features from device-specific characteristics through cross-batch recombination of content and style components during segmentation and reconstruction. This process preserves structural consistency across imaging conditions while discarding scanner-induced artifacts. To further bolster resilience, two complementary innovations are integrated: Expansion Mask Attention (EMA) prioritizes anatomical boundaries in low-contrast regions, while Style Augmentation (SA) emulates diverse acquisition parameters by generating synthetic style variations. Validation across optic disc/cup and prostate segmentation tasks demonstrates the framework's efficacy. In multi-to-single-center generalization, UniDDG achieved Dice scores of 84.43% (optic disc and optic cup) and 86.96% (prostate), and achieved that of 88.91% and 88.56% respectively for single-center generalization—surpassing several existing state-of-the-art methods in Dice scores. These results highlight its capacity to mitigate domain discrepancies without architectural modifications, offering a scalable solution for real-world clinical heterogeneity.

## 1. Introduction

Deep learning has significantly advanced the field of medical image segmentation, enabling automated and highly accurate delineation of anatomical structures across various imaging modalities. Notable breakthroughs, such as U-Net [1] and its derivatives, have become foundational architectures for tasks like organ segmentation, tumor detection, and other critical medical applications [2, 3]. These models, however, typically rely on a key assumption: that the training data and test data share the same underlying distribution. This assumption, known as the independent and identically distributed (i.i.d.) hypothesis, has underpinned much of the success in medical image segmentation. In reality, however, training data and test data often come from different distributions, leading to a domain shift that negatively impacts model performance when deployed in clinical practice [4-6]. As a result, models that perform well on training datasets may struggle to achieve the same level of accuracy on new, unseen data, underscoring the need to address domain shift in medical image segmentation [7].

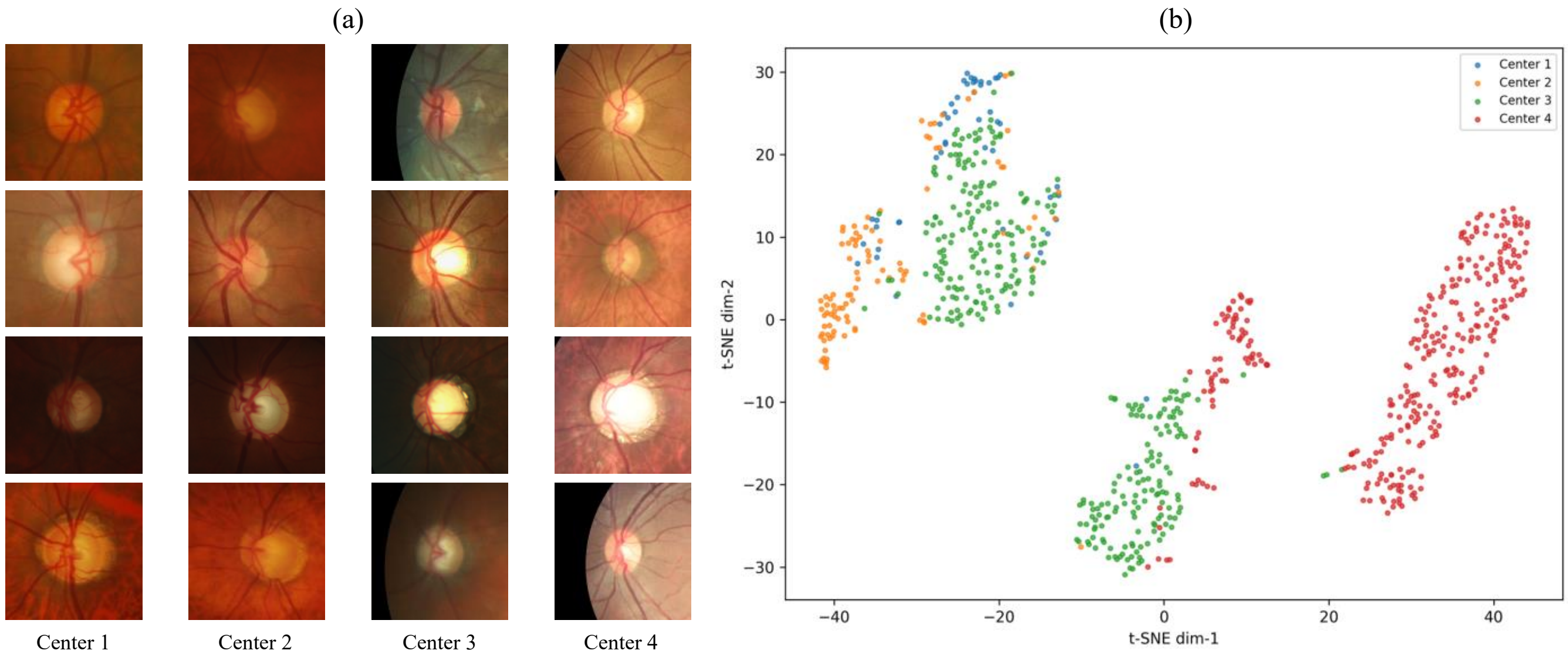

Figure 1 Illustration of image differences between centers and within centers. (a) Examples of highly dissimilar images within the same center and similar images across different centers; (b) t-SNE distribution visualization of data from all centers.

To address the issue of reduced performance in trained networks across different domains, researchers have explored a range of strategies for medical image analysis. One common approach is transfer learning, specifically through fine-tuning, adjusting network weights to fit the unique characteristics of a new domain. However, fine-tuning often requires additional labeled data from the target domain, which presents challenges due to the high cost and effort required for medical image annotation. In response, various unsupervised domain adaptive techniques have been developed to reduce the distribution gap between source and target domains without the need for labeling of the latter[8, 9]. By facilitating knowledge transfer across domains, this approach has been shown to be effective in medical image segmentation [10]. However, as effective as domain adaptation (DA) is, it faces several major challenges[11]. First, DA usually requires access to target domain data during training, but this is not always available in the real world due to privacy concerns, lack of labeled data, or inability to get data from unseen domains [12]. Secondly, DA methods usually lead to catastrophic forgetting, that is, the performance of the model on the source domain deteriorates after adapting to the new target domain [13]. This phenomenon is particularly problematic in the healthcare environment, where the model must maintain high performance in all domains, including the original source domain.

On the other hand, the domain generalization (DG) approach addresses these challenges by enhancing the model's ability to generalize to unknown domains without the need to access target domain data during training and without the need for catastrophic forgetting [14-18]. While DG techniques have successfully closed the performance gap between source and target domains in natural image classification[19-21], their application in medical image segmentation presents unique challenges due to the complex and structured predictive nature of segmentation tasks[22]. In the field of medical imaging, CNN-based approaches explore data enhancement to improve generalization; However, they often rely on single-source domain training, which limits their ability to capture complex relationships between multiple domains that could improve the model's performance on previously unseen target data [23].

To address these limitations, recent DG methods have taken two primary directions: image-based and feature-based approaches. Image-based DG methods, such as BigAug, apply extensive data augmentation by stacking transformations to mimic domain shifts, aiming to bridge the domain gap through simulation [24]. However, the effectiveness of these augmentations often depends on data-specific configurations, which limits their generalizability across diverse medical imaging environments. Feature-based DG methods, by contrast, emphasize learning domain-invariant representations to capture core features that remain consistent across domains. Domain-oriented feature embedding, for example, uses knowledge from multi-site datasets to make semantic features more domain-discriminative [12], while techniques like Domain Composition and Attention Network (DCA-Net) use a representation bank and attention modules to dynamically adapt features based on domain context [25]. Although these methods benefit from domain-specific knowledge pools, their reliance on predefined representation banks restricts generalization across a broader range of domains.

In recent developments, feature disentanglement has been explored to separate domain-invariant content from domain-specific styles, which enables models to learn flexible representations across modalities. However, traditional GAN-based disentanglement methods often require multiple encoders and discriminators for each known domain, resulting in a complex training process and limited scalability to unseen domains. A CDDSA framework addresses these limitations with introducing a contrastive learning approach that disentangles domain-invariant content while augmenting style representations to increase model adaptability [13]. By reducing dependence on complex GAN architectures and extensive domain-specific encoders, CDDSA enhances scalability and allows for improved generalization to unseen domains, marking a significant step forward in the robustness of medical image segmentation across diverse clinical settings.

Although these methods are very promising, they often treat different centers as different domains when facing multi center scenarios in the source domain, ignoring the fact that significant domain changes may occur within one center. In some cases—particularly after observing various multi-center datasets—the variability within a center, due to different scanner models or imaging protocols, may result in domain shifts even greater than those observed between centers. As shown in Figure 1a, there are significant differences in the images within each center, and in some cases, images from different centers may exhibit similarities. To quantitatively verify these complex domain shifts, we visualized the distribution of all data from all centers using t-SNE, as shown in Figure 1b. The visualizations of raw data from different centers reveals that the mean intra-center distance is comparable to the mean inter-center distance, and the center clusters overlap, confirming the existence of domain discrepancies. However, the visualization also highlights a crucial phenomenon: domain clusters are not strictly separable. There is substantial overlap where images from different centers share similar stylistic features. *This phenomenon makes it unreliable to classify images from a single center into one domain and to categorize images from different centers into distinct domains*. Based on this, we believe that rethinking the current domain partitioning scheme is highly necessary to improve the model's generalization ability.

A more granular scanner- or protocol-based domain partition would require complete acquisition metadata for every image, which are often unavailable and make explicit domain identification impractical.

The central research gap lies in how domain-specific style is defined and constrained in disentanglement-based DG. Image-level augmentation and sample-wise feature-statistics mixing can diversify training appearances [20, 24], but they do not resolve whether center identity provides a reliable domain partition for learning style representations. In CDDSA[13], style codes from the same predefined domain are encouraged to be compact, whereas those from different domains are encouraged to be divergent. This objective assumes that center membership provides a reliable partition of acquisition style. However, the substantial intra-center variation and inter-center overlap observed in Figure 1 indicate that this assumption may impose inaccurate relationships among style representations. The unresolved problem is therefore to learn and recombine image-specific style representations without using center labels to define those relationships. UniDDG addresses this gap by operationalizing the OIOD hypothesis within a disentanglement framework that does not use center labels to define style relationships.

This paper proposes a new perspective of domain generalization by redefining the concept of domain. Instead of grouping images according to scanner or central features, we treat each image as a unique domain, assuming that each image has its own domain shift and style variations. This approach, which we call the "One Image as One Domain" (OIOD) hypothesis, redefines the concept of domain division in medical image segmentation. Unlike traditional data augmentation techniques that merely increase data diversity, OIOD fundamentally restructures the learning paradigm by equating the domain space with the sample space. This structural shift eliminates the dependency on domain labels, allowing the model to internalize domain invariance at the instance level rather than the center level. Based on this assumption, we develop a unified disentanglement-based domain generalization (UniDDG) framework for medical image segmentation that can be flexibly adapted to scenarios involving single and multi-source domains. The framework enhances the ability to generalize to various unknown domains without the need for explicit domain labels. In addition, it maintains a fixed computational complexity and workload that does not increase with the number of source domains, and it simplifies the training process by not relying on Gans. Our work contributions are summarized below:

(i) We explicitly formulate each training image as an independent domain for domain-generalizable medical image

segmentation and to operationalize this OIOD hypothesis through a disentanglement objective that does not use center identity to organize style representations. This formulation enables UniDDG to learn and recombine image-specific style representations in both single-source and multi-source settings without explicit domain labels.

(ii) We propose the UniDDG framework, a domain generalization method based on unified disentanglement that is suitable for both single-source and multi-source scenarios, ensuring consistent computational complexity and scalability regardless of the number of source domains.

(iii) We introduce Expansion Mask Attention (EMA), which applies an extended mask to focus the target area and its surroundings during reconstruction, enhancing boundary protection and improving segmentation accuracy.

(iv) We developed Style Augmentation (SA), which generates random style code to simulate different image styles, improving the robustness of the model to domain variations and enhancing its generalization ability.

(v) Extensive experiments on two segmentation tasks (optic disc and cup segmentation and prostate segmentation) show that our proposed framework has superior performance and adaptability compared to the most advanced methods.

## 2. Related work

### 2.1. Domain adaptation for image segmentation

Domain adaptation (DA) has become a key method for solving domain shift problems in medical image segmentation, especially when training and testing data come from different distributions[26]. The traditional methods in DA, such as fine-tuning pre trained networks on new domain data, aim to minimize the distribution gap between the source and target domains. However, this typically requires labeled data from the target domain, which is difficult to obtain in medical environments and costly. To address this issue, unsupervised domain adaptation (UDA) is proposed, in which the model learns to adapt to the target domain without requiring any labeled data. Various unsupervised domain adaptation techniques have been proposed to address cross-domain discrepancies. These include feature-level alignment [27-35], pixel-level alignment [36-39], self-learning approaches [40-43] , as well as hybrid methods that combine these strategies [4, 10, 44, 45].

Current research is limited to two domains (a source domain and a target domain), failing to fully leverage labeled data from many other centers. In the field of image segmentation, two representative works have demonstrated the advantages of multi-source unsupervised domain adaptation over single-source methods when dealing with multi-centered source domain data [46, 47]. Segmentation accuracy in multi-source domains is significantly higher than in single-source domains. Zhao et al. [46] combined pixel-level alignment with feature distribution alignment, using GANs to transform two source domain images into target domain style images, which are then merged into a single source domain. They applied adversarial learning to implicitly align feature distributions between the source and target domains. When extending this method to more source domains, the number of adversarial training iterations must increase accordingly. Overuse of adversarial training not only increases model complexity but may also prevent the model from converging to a Nash equilibrium [48, 49]. He et al. [47] integrated pixel-level alignment with self-supervised learning, converting RGB images into LAB color space, and explicitly shifting pixel distributions to transform source domain images into the target domain style. They further improved model performance by using feature distribution differences or pseudo-labels from two model outputs in a mutual learning framework. The number of segmentation networks in this framework equals the number of source domains, and when scaled to more source domains, substantial computational resources are required, particularly for three-dimensional image inputs. Recent advancements have further expanded the scope of domain adaptation through more sophisticated alignment strategies. For instance, TriLA [50] introduced a triple-level alignment mechanism to enforce feature consistency for joint optic disc and cup segmentation, while reconstruction-driven approaches [51] utilize dynamic refinement to improve unsupervised adaptation. Furthermore, the emergence of foundation models has inspired prompt-learning frameworks [52], which leverage pre-trained knowledge for source-free adaptation. However, these methods often still operate under traditional domain definitions or require explicit adaptation phases.

### 2.2. Domain generalization for image segmentation

Although domain adaptation (DA) aims to use data from that domain to adapt the model to the target domain, it requires access to the target domain data during training and may lead to catastrophic forgetting of the source domain. To overcome these challenges, domain generalization (DG) technology has become a promising alternative[53]. Domain generalization focuses on enhancing the model's ability to generalize to unknown domains without the need to use target domain data during training, otherwise it may reduce the performance of the source domain[54]. In the context of image segmentation, DG aims to improve the robustness of models in various domains, including unseen domains, by utilizing data augmentation and other strategies. Domain adaptation uses target-domain data during training, whereas domain generalization (DG) learns from source domains alone and is evaluated on unseen target domains[53, 54]. In medical image segmentation, DG therefore seeks representations that remain robust to acquisition- and center-related shifts without accessing target images during training.

For clarity, we categorize the reviewed DG methods according to their primary mechanism for handling domain shift: (1) data augmentation and style randomization; (2) domain-invariant feature learning through semantic constraints, normalization, or feature calibration; (3) source-aware and multi-source representation learning; (4) probabilistic style modeling and content-style disentanglement; and (5) foundation-model-based prompting. Because some methods combine multiple mechanisms, this categorization follows the component that primarily contributes to cross-domain generalization. It also provides a consistent basis for positioning UniDDG relative to the closest disentanglement-based approaches.

The first direction used for image segmentation in DG is related to data enhancement, such as Manifold Mixup [55] and CutMix [56], which create new training samples by interpolating or mixing images in a feature space. These methods have been shown to be effective in generating domain invariant representations, enhancing the robustness of the model in unknown environments. Zhang et al. [24] proposed the idea of deep stacked transform (BigAug), extended the deep learning model to medical image segmentation, and used the transformation of simulated domain shift to improve the model's generalization ability to unknown data sets. These methods rely heavily on the premise that augmentation can simulate various variations in the data, allowing it to effectively improve the model's adaptability to new domains without the need for labeled data from those domains. Instance normalization and related style-manipulation methods operate on individual feature statistics. MixStyle [20], for example, mixes instance-level statistics across samples to generate new styles, while BigAug [24], Manifold Mixup [55], and CutMix [56] diversify training through image- or feature-level transformations. In these approaches, the sample is the unit on which an operation is applied. OIOD makes a different formulation-level choice: the image is the domain unit used to organize domain-specific style within a disentanglement objective. Accordingly, UniDDG learns an explicit style representation for each image and recombines content and style across pooled training samples without center-based grouping. The distinguishing contribution of OIOD therefore lies in the domain formulation that governs style disentanglement, rather than in applying an instance-level normalization, randomization, or augmentation operation.

Another work focuses on semantic perception domain generalization, which attempts to improve segmentation by using semantic information to guide the model's generalization ability. Peng et al. [57] proposed a semantic-aware domain generalized segmentation approach that learns robust representations across different domains, ensuring that the model's predictions remain consistent even in the presence of domain shifts. Other approaches, such as those proposed by Choi et al. [58], enhance model robustness in segmentation tasks by introducing selective whitening to reduce domain-specific variations, achieving improved performance on unseen domains. However, many of these methods still rely on single-source domain data and fail to fully leverage multi-source domain data, which could provide more comprehensive insights into the generalization process.

When extending DG techniques to multi-source domains, several approaches have been proposed, each with its own strengths and limitations. For example, Dofe [12] employs domain-oriented feature embedding for fundus image segmentation; however, its reliance on fixed embeddings may not fully capture the rich variability inherent in real-world data. Similarly, D-CAC [59] adapts convolutional layers to reconcile features from different domains, but this explicit adaptation can be sensitive to overlapping domain characteristics and may not scale efficiently as the number of source domains increases. Feddg [60] leverages federated learning to address domain variability across sources, yet its decentralized training introduces additional complexity that can hinder robust feature alignment. Among more recent methods, TriD [61] uses domain randomization to generate diverse imaging conditions, effectively extracting domain-invariant features. Still, its reliance on simulated variations may not capture the complex variations that exist in real-world data sets. At the same time, DFQ [62] decoupled semantic and spatial feature extraction to mitigate overfitting, but this decoupling increased computational complexity and posed challenges for seamlessly integrating extracted features. DAPSAM [63] introduces a domain-adaptive prompt learning framework that fine-tunes the Segment Anything Model (SAM) for single-source generalization. MixStyleFlow [64] utilizes normalizing flows to explicitly model and sample domain feature styles, expanding the diversity of training data. Additionally, Aegis [65] proposes a dual attention-guided feature calibration module to mitigate cross-domain feature misalignment.

These recent methods represent three distinct directions in generalizable medical image segmentation. DAPSAM[63] transfers the pretrained representation of SAM through domain-adaptive prototype prompting and is designed primarily for single-source generalization. MixStyleFlow [64] explicitly models the probability distribution of feature styles using normalizing flows, whereas Aegis [65] focuses on reducing cross-domain feature misalignment through attention-guided calibration. UniDDG differs from these approaches at the domain-formulation level: it treats each image as the unit of style variation and constrains global style exchange through content, style, and prediction consistency. This formulation does not require center labels and is applied without architectural changes in both single-source and multi-source settings.

Style-oriented DG methods modify or sample image appearance at the input, feature, or latent level. TriD[61] uses domain randomization to synthesize diverse imaging conditions, MixStyleFlow[64] models and samples feature-style distributions, and CDDSA [13] recombines latent style representations within a disentanglement framework. Meta-learning-based DG constructs simulated generalization episodes from the available source data; Feddg [60] incorporates episodic learning into a federated medical segmentation setting. Foundation-model-based DG transfers pretrained representations to unseen-domain segmentation. DAPSAM [63] follows this direction by adapting the Segment Anything Model through domain-adaptive prototype prompting. VAE-based style encoding represents each image through the parameters of a latent distribution [66]. UniDDG uses this probabilistic encoding for image-specific style representation and constrains the resulting style codes through reconstruction and consistency objectives. These method families differ in how they expose the segmenter to domain variation: appearance randomization expands the observed style support, episodic learning simulates source-to-unseen-domain transfer, foundation-model prompting adapts pretrained representations, and VAE-based encoding models variation in an explicit latent style space.

In general, these approaches - from fixed embedding and adaptive convolution methods to simulation-based and decoupling strategies - face scalability, high computational overhead, and difficulty capturing subtle cross-domain variations. This highlights the need for a more robust and efficient approach to the challenges of multi-source domain generalization. To address these limitations, we propose a new approach based on style and content disentanglement, as explored in the CDDSA framework [13]. Our method treats each image as a unique domain, thereby eliminating the need for explicit domain labels while maintaining a fixed computational complexity regardless of the number of source domains. With a streamlined architecture comprising a single content encoder, a style encoder, a segmentation network head, and a decoder, UniDDG maintains a fixed model size and per-sample computation graph regardless of the number of source domains. Moreover, the style encoder and decoder are discarded after training, leaving only the content encoder and segmentation head for inference.

CDDSA [13] is the closest architectural predecessor to UniDDG. Both frameworks use a shared anatomical/content encoder, style encoder, segmentation head, and reconstruction decoder, and neither instantiates a separate network for each source domain. The principal distinction therefore lies in the domain formulation and training objective. CDDSA uses predefined domain labels to construct same-domain positive pairs and different-domain negative pairs, explicitly encouraging within-domain compactness and between-domain separation in the style space. It also applies KL regularization toward a unit Gaussian prior and generates augmented styles through random linear combinations of a source-domain style-code bank. In contrast, UniDDG treats each image as the finest observable unit of style variation and does not use center labels to construct style groups or exchange pairs. Image-specific style codes are globally reassigned, and the reconstructed images are constrained at the content, style, and prediction levels. UniDDG further applies EMA to emphasize task-relevant regions during reconstruction. Consequently, CDDSA encodes center membership as a supervisory relationship between style codes, whereas UniDDG allows appearance similarity to be learned independently of center membership. Both methods maintain a shared model architecture as the number of source domains increases, but UniDDG removes label-conditioned contrastive grouping and applies the same objective to both single-source and multi-source settings.

At the representation level, CDDSA [13] and UniDDG share a content-style decomposition architecture. Both methods encode anatomical information as a spatial content representation, represent image appearance using a latent style code, and inject the style code into the reconstruction decoder through Adaptive Instance Normalization (AdaIN). The style encoder of UniDDG follows the VAE parameterization [66] by predicting the mean and variance of an image-conditioned latent distribution. AdaIN converts the sampled style code into channel-wise scaling and bias parameters that modulate the content features during reconstruction. UniDDG contributes a different domain formulation and training objective on this architectural foundation. CDDSA uses center labels for contrastive grouping of style codes, applies KL regularization toward a shared prior, and constructs augmented styles from a source-domain style bank. UniDDG represents style at the image level, removes center-conditioned grouping and KL prior matching, globally reassigns image-specific style codes, and constrains the exchanged representations through content, style, and prediction consistency. The methodological distinction lies in how the latent style space is defined, exchanged, and supervised rather than in the use of VAE or AdaIN operations alone.

The absence of source-center labels can correspond to different modeling assumptions. Domain-agnostic learning aims to learn representations that remain applicable across domains without conditioning predictions on a predefined domain identity [17]. Domain-label-free DG removes observed center annotations from the training input or objective, but the absence of such annotations does not itself determine how domain-specific variation is represented. Instance-level representation learning uses individual samples as optimization units, whereas sample identity does not necessarily define the style factor from which anatomical content should be separated. OIOD makes a more specific modeling choice: each image defines an image-conditioned style representation, style codes are exchanged globally without constructing center-based or pseudo-domain groups, and the reconstructed images are re-encoded under content, style, and prediction consistency constraints. OIOD therefore uses image identity to define and supervise domain-specific style variation rather than only omitting domain labels or applying an instance-wise objective.

## 3. Method

### 3.1. The unified disentanglement-based domain generalization framework

Multi-center-based domain generalization for medical image segmentation is generally defined as enhancing the ability of a segmenter $S$, trained on datasets $\{D_1, \ldots, D_K\}$ from $K$ labeled centers, to generalize effectively to an additional dataset $D_{K+1}$ from an unobserved target center. However, this paper introduces a new perspective on domain partitioning, treating each image as an individual domain with its own distinct style, even if images come from the same center. As a result, data from multiple source centers can be mixed together without the need to classify images by center-based domains. Specifically, the training data is represented as $\{x_1, \ldots, x_t\}$, where $x$ denotes an individual image, and $t$ represents the total number of images across all source domains.

The proposed unified disentanglement-based domain generalization (UniDDG) framework is illustrated in Figure 2. All source domain images are mixed in a data pool, from which a random batch of $n$ images $\{x_1, \ldots, x_n\}$ is selected for training. This batch of images is then fed to the content encoder *Econ* and the style encoder *Esty* to perform feature disentanglement, resulting in the corresponding content representation $\{c_1, \ldots, c_n\}$ and style code $\{s_1, \ldots, s_n\}$. Subsequent operations involve exchanging and combining the content representation and style code in the batch for segmentation, reconstruction, and further disentanglement. Under OIOD, each training image is treated as the finest observable unit at which style variation is modeled. This image-level domain definition does not require every image to arise from a statistically independent distribution or force different images into isolated style clusters. Instead, it avoids imposing potentially inaccurate style relationships solely according to center membership. Images with similar appearances are still allowed to learn similar style codes, regardless of whether they originate from the same center. Specifically, each image is decomposed into an anatomical representation $c_i$and an image-specific style code $s_i$. The style codes are globally reassigned across the batch without using center labels, thereby exposing each anatomical representation to both intra-center and inter-center appearance variations. These exchanges are not treated as valid merely because they are randomly generated. The content consistency loss requires the reconstructed image to preserve $c_i$, the style consistency loss requires it to retain the reassigned style, and the segmentation consistency loss requires its anatomical prediction to remain unchanged. Consequently, appearance changes that corrupt anatomical information are penalized rather than accepted as meaningful domain variation. OIOD therefore implements a finer-grained invariance objective over the empirical image-level style support, while the consistency constraints control the potential noise introduced by global style exchange. The following steps outline how to use content presentation and style code:

The distinction between OIOD and sample-wise augmentation is implemented at the objective level. A conventional sample-wise augmentation transforms an input image or its feature statistics and then applies the task loss to the transformed sample. In UniDDG, each image $x_i$ is instead decomposed into an anatomical representation $c_i$ and an image-specific style code $s_i$. The style code is globally reassigned to another content representation, producing a reconstructed image $D(c_i, s_{\pi(i)})$, which is then re-encoded. The content, style, and segmentation consistency losses jointly require the reconstructed image to preserve the anatomy of $x_i$ while expressing the reassigned style $s_{\pi(i)}$. Center labels are not used to select exchange pairs or to determine which style representations should be grouped. Therefore, OIOD changes not only how an individual sample is transformed, but also how domain-specific information is defined, exchanged, and constrained throughout training.

(i) The content to be obtained represents $\{c_1, \dots, c_n\}$ is entered into the segmenter $S$ to obtain the predicted split result $\{y'_1, \dots, y'_n\}$. By calculating the difference between these predictions with ground truth values $\{y_1, \dots, y_n\}$ and the resulting segmentation loss $\mathcal{L}_{seg}$.

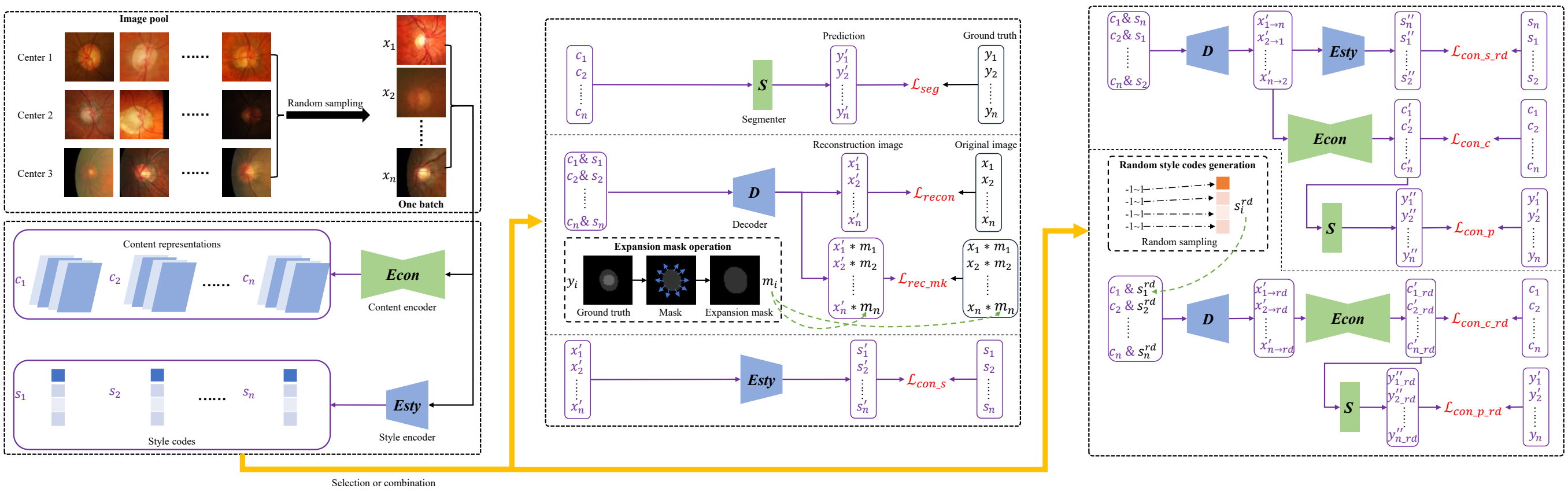

Figure 2 Overview of the proposed unified disentanglement-based domain generalization framework.

(ii) The content representations $\{c_1, \dots, c_n\}$ and style codes $\{s_1, \dots, s_n\}$ are paired and input into a decoder $D$ to reconstruct the original images $\{x'_1, \dots, x'_n\}$. By calculating the discrepancy between the reconstructed images and the originals $\{x_1, \dots, x_n\}$, the reconstruction loss $\mathcal{L}_{recon}$ is derived. To focus $D$ more on the segmentation region and its surrounding areas during reconstruction, an expansion mask is introduced. Specifically, the ground truth $y_i$ is converted into a binary mask, and the boundary of the target area in the mask is expanded outward by a certain distance to generate the expansion mask $m_i$. This mask is then multiplied with both the reconstructed and original images to calculate their discrepancy, yielding the masked reconstruction loss $\mathcal{L}_{re_mk}$.

(iii) The reconstructed images $\{x'_1, \dots, x'_n\}$ are re-disentangled by inputting them into $Esty$ to extract new style codes $\{s'_1, \dots, s'_n\}$. The discrepancy between these new codes and the original style codes $\{s_1, \dots, s_n\}$ is used to define the style consistency loss $\mathcal{L}_{con_s}$.

(iv) The content representations $\{c_1, \dots, c_n\}$ and style codes $\{s_1, \dots, s_n\}$ are randomly paired and input into $D$ to generate reconstructed images with exchanged styles $\{x'_{1\to n}, \dots, x'_{n\to 2}\}$. These reconstructed images, in which the styles have been swapped, are then passed through $Econ$ and $Esty$ to re-disentangle them, obtaining content representations $\{c'_1, \dots, c'_n\}$ and style codes $\{s''_n, \dots, s''_2\}$. By calculating the discrepancy between these representations and the content representations $\{c_1, \dots, c_n\}$ and style codes $\{s_1, \dots, s_n\}$, content and style consistency losses $\mathcal{L}_{con_c}$ and $\mathcal{L}_{con_{s_rd}}$ are obtained. The content representations $\{c'_1, \dots, c'_n\}$ are also fed into $S$ to get segmentation predictions $\{y''_1, \dots, y''_n\}$. By calculating the difference between these predictions and the initial segmentation results $\{y'_1, \dots, y'_n\}$, resulting in a segmentation consistency loss $\mathcal{L}_{con_p}$. It is important to note that the Style Encoder ($E_{sty}$) and Decoder (D) are auxiliary modules designed solely to enforce feature disentanglement and domain invariance during the training phase. During the inference phase, these modules are discarded, and only the Content Encoder ($E_{con}$) and Segmenter (S) are retained. Consequently, the deployed model incurs no additional computational overhead in terms of parameters or inference latency compared to the backbone network.

(v) To enhance the robustness of the segmenter and content encoder against variations in image styles, additional styles $s_i^{rd}$ are randomly generated named style augmentation. These new style codes are combined with the initial content representations $\{c_1, \dots, c_n\}$ and input into $D$ to produce randomly styled reconstructed images $\{x'_{1\to rd}, \dots, x'_{n\to rd}\}$. These reconstructions are then input into $Econ$ to obtain content representations $\{c'_{1_rd}, \dots, c'_{n_rd}\}$. The discrepancy between these representations and the initial content representations yields the content consistency loss $\mathcal{L}_{con_c_rd}$. Furthermore, $\{c'_{1_rd}, \dots, c'_{n_rd}\}$ is also input into $S$, generating predicted segmentation results $\{y''_{1_rd}, \dots, y''_{n_rd}\}$. By calculating the discrepancy between these predictions and the initial segmentation results $\{y'_1, \dots, y'_n\}$, the segmentation consistency loss $\mathcal{L}_{con_p_rd}$ is obtained.

### 3.2. Content encoder and segmenter

We use a modified U-Net [1] as the backbone to implement $Econ$, adjusting the final layer's output channel to $R$ and using tanh activation. For input image $x_i$ with height $H$ and width $W$, the output $c_i \in [-1,1]^{H\times W\times R}$ is structured so each channel captures specific anatomical details. Unlike SDNet [67], which restricts $c_i$ to binary values, our approach uses tanh activation $tanh(x) = \frac{e^x - e^{-x}}{e^x + e^{-x}}$ to retain detailed structures, enhancing image reconstruction and style augmentation. Specifically, the Content Encoder is designed with a channel progression of [16, 32, 64, 128, 256] corresponding to five resolution scales, ensuring sufficient capacity to capture hierarchical anatomical features. The anatomical representation extraction is:

$$c_i = E_{con}(x_i) \quad (1)$$

This representation $c_i$ is input to a segmentation network $S$, shown as in Figure 3, producing a probability map $y'_i = S(c_i)$. Let $y_i$ denotes the ground truth, the segmenter $S$ is optimized by minimizing the segmentation loss $\mathcal{L}_{seg}$ between $y'_i$ and $y_i$. Dice coefficient is employed as the segmentation loss in this study [68]:

$$\mathcal{L}_{seg} = \frac{\sum_{i=1}^{N} 2y'_i y_i}{\sum_{i=1}^{N} y_i y_i + \sum_{i=1}^{N} y'_i y'_i} \quad (2)$$

where $N$ represents the number of the inputs.

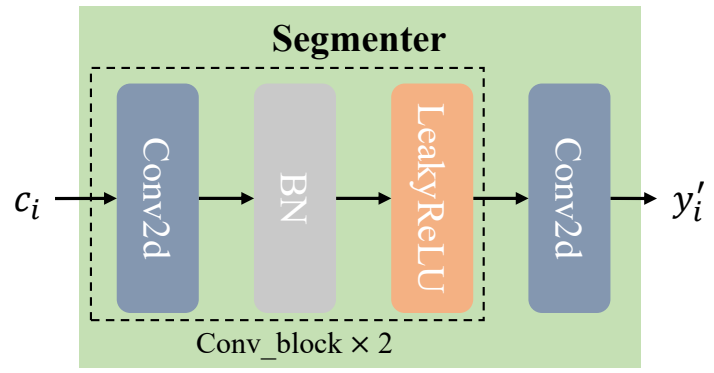

Figure 3 Structure of the segmentation network.

### 3.3. Style encoder

In our approach, the domain-specific modality representations are generated by a style encoder $Esty$, implemented as the encoder part of a Variational Autoencoder (VAE) [66], as shown in Figure 4. The VAE typically learns a low-dimensional latent space, where the learned latent representation approximates an isotropic multivariate Gaussian distribution $p(z) = \mathcal{N}(0,1)$. For each input $x_i$, $Esty$ predicts the mean $\mu_i$ and variance $\sigma_i$ of the latent distribution $z \in \mathbb{R}^{1\times Z}$, where $Z$ is the length of the latent code. The encoder backbone consists of four convolutional blocks with down-sampling, followed by a global average pooling layer and two parallel fully connected layers that map the features to the mean $\mu_i$ and variance $\sigma_i$ vectors of the latent space. The style code $s_i$ for an input image $x_i$ is then sampled from the predicted distribution defined by the mean $\mu_i$ and variance $\sigma_i$.

Specifically, we omit the $KL$ divergence that would conventionally regularize each image-conditioned distribution $q(z \mid \mu_i \sigma_i)$toward the same unit Gaussian prior $p(z)$. In UniDDG, the style encoder is used to represent image-conditioned appearance rather than to support unconditional generation from a shared prior. Enforcing prior matching would therefore discourage the image-specific deviations required for style exchange. The latent representation remains constrained by its low dimensionality ($Z = 48$) and by the two style consistency losses, which require reconstructed images to recover their original or reassigned style codes. An unstable or unrecoverable style code directly increases these consistency penalties, while the reconstruction, content, and prediction consistency losses further require the generated image to remain valid and anatomically consistent. Thus, UniDDG regularizes the style representation according to its functional role in reconstruction and disentanglement instead of its distance from a common prior.

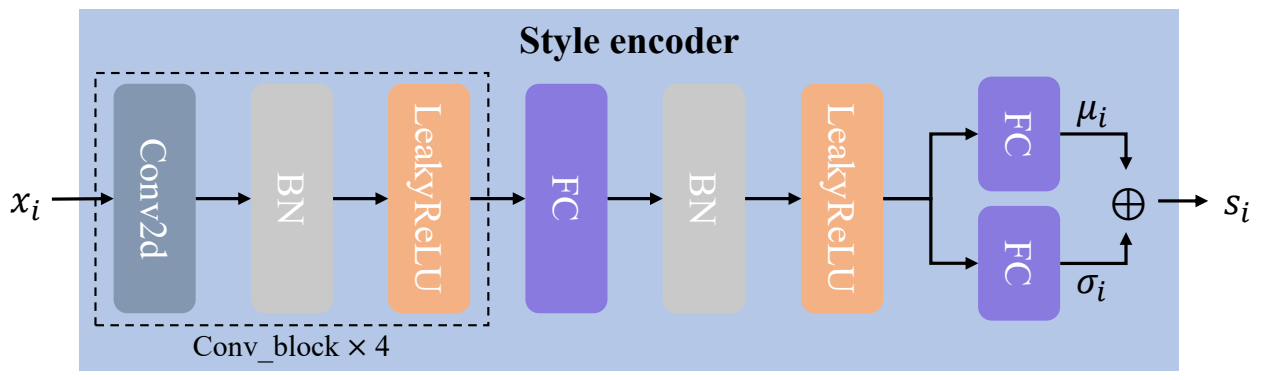

Figure 4 Structure of the style encoder[13].

### 3.4. Decoder

In this study, we adopt the same reconstruction decoder $D$ structure as in reference [13]. Figure 5 presents the architecture of $D$ , which produces a reconstructed image $x_i'$ by combining two separate representations, $c_i$ for content and $s_i$ for style. These two representations collaborate in a "repainting" role: $c_i$ defines the structural details, while $s_i$ supplies stylistic features across the image [69].

To generate the final image, the decoder processes $c_i$ through three repainting blocks for style adjustments and one convolution layer. In each repainting block, the adaptive scaling $\alpha$ and bias $\beta$ parameters that are used in Adaptive Instance Normalization (AdaIN) can be predicted with the guidance of the style code $s_i$. Specifically, AdaIN individually transforms each channel of the feature map outputted by the previous convolution layer in each repainting block, adjusting style through calculated $\alpha$ and $\beta$ values for every channel.

The decoder thus progressively layers style information over the content structure, adapting from coarse to refined details. The final output image $x_i'$ can be expressed as:

$$x_i' = D(c_i, s_i) \tag{3}$$

Because $c_i$ and $s_i$ originate from $x_i$, the reconstructed image $x_i'$ should approximate the original input as closely as possible. To achieve this, we apply a reconstruction loss based on $L1$ loss function for its robustness to outliers:

$$\mathcal{L}_{recon} = \frac{1}{N}\sum_{i=1}^{N}|x_i' - x_i| \tag{4}$$

where $N$ represents the number of the inputs.

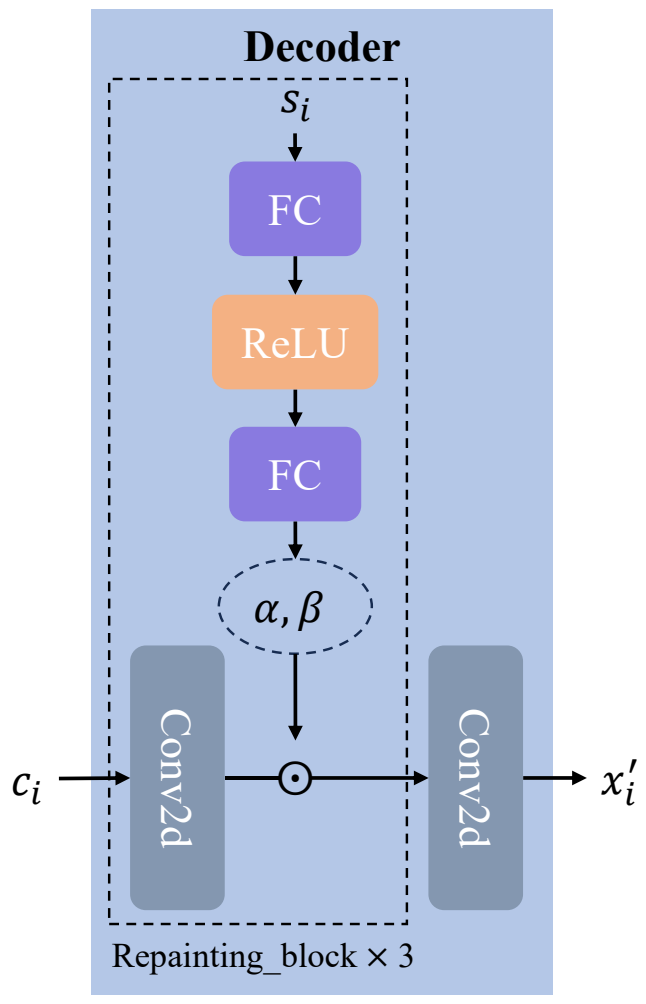

Figure 5 Structure of the reconstruction decoder[13].

### 3.5. Expansion mask attention for reconstruction

In the previous section, the reconstruction loss was computed based on the difference between the entire reconstructed image and the original, controlling the overall quality of the image restoration. Since it is challenging for the decoder to perfectly reconstruct an image identical to the original (i.e., the reconstruction loss cannot be zero), we want the model to focus more on the target segmentation area and its surrounding regions. This is because the primary objective of segmentation is to identify the boundaries of the target object, and for accurate reconstruction, the boundaries must be precisely captured. As a result, the reconstruction should emphasize not only the segmented target area but also its surrounding context. We specifically chose this region-based attention mechanism over explicit edge-based losses (e.g., Sobel gradients) because medical images often exhibit low contrast and high noise, making pixel-level edge detection unstable. Unlike gradient-based methods that can be misled by artifacts, EMA defines a broader "boundary zone." This zone provides a more robust supervisory signal for structural learning than noisy edge maps, effectively preventing large background areas from dominating the reconstruction loss while maintaining resilience to local intensity variations. To guide this process, we introduced expansion mask attention (EMA), which specifies the region where more precise reconstruction is needed, ensuring the model prioritizes accurate restoration in these key areas.

The process of obtaining the expansion mask is illustrated in Figure 2. First, the ground truth mask is binarized, with pixel values set to 0 for the background and 1 for the target object. The target region (pixels with a value of 1) is then dilated, slightly extending the mask beyond the object boundary. The expansion mask $m_i$ is subsequently applied to both the reconstructed and original images via element-wise multiplication, and a reconstruction loss $\mathcal{L}_{rec_mk}$ is calculated between the two, $x_i' * m_i$ and $x_i * m_i$. This process directs the decoder to focus on the regions covered by the expansion mask, thereby enhancing the preservation of boundary details between the segmented object and the surrounding background. The masked reconstruction loss $\mathcal{L}_{rec_mk}$ which is based on based on $L1$ loss can be defined as:

$$\mathcal{L}_{rec_mk} = \frac{1}{N}\sum_{i=1}^{N}|x_i' * m_i - x_i * m_i| \tag{5}$$

### 3.6. Style augmentation

To enhance the robustness of the content encoder and segmenter against image style variations, we adopt a style augmentation (SA) approach. As shown in Figure 2, we first generate random style code by sampling each element from a uniform distribution in the range of -1 to 1. This random style code $s_i^{rd}$ is then combined with the content representation $c_i$ obtained from the initial decoupling and fed into the decoder $D$ to produce a style-augmented reconstruction image $x_{i\to rd}'$. This image is subsequently passed through the content encoder $Econ$, and the resulting content representation $c_{i\to rd}'$ is constrained to match the original content representation $c_i$ via a consistency loss $\mathcal{L}_{con_c_rd}$. $c_{i\to rd}'$ is then input into the segmenter, where the predicted segmentation output $y_{i\to rd}''$ is required to align with the original segmentation result $y_i'$, again governed by a consistency loss $\mathcal{L}_{con_p_rd}$. The above two consistency losses are based on based on $L1$ loss:

$$\mathcal{L}_{con_c_rd} = \frac{1}{N}\sum_{i=1}^{N}|c_{i\to rd}' - c_i| \tag{6}$$

$$\mathcal{L}_{con_p_rd} = \frac{1}{N}\sum_{i=1}^{N}|y_{i\to rd}'' - y_i'| \tag{7}$$

The U(-1,1) range applies to the random codes drawn by SA; it is not a constraint on the codes produced by the style encoder.

### 3.7. Overall loss function

The UniDDG framework incorporates a total of nine loss functions. Five of these are defined in previous sections, while the remaining four—$\mathcal{L}_{con_s}$, $\mathcal{L}_{con_c}$, $\mathcal{L}_{con_s_rd}$, and $\mathcal{L}_{con_p}$—are consistency loss functions based on L1 loss, and thus their detailed formulations are omitted for brevity. The overall loss function of the UniDDG framework is as follows:

$$\mathcal{L}_{all} = \lambda_1\mathcal{L}_{seg} + \lambda_2\mathcal{L}_{recon} + \lambda_3\mathcal{L}_{rec_{mk}} + \lambda_4\mathcal{L}_{con_s} + \lambda_5\mathcal{L}_{con_{s_{rd}}}$$
$$+\lambda_6\mathcal{L}_{con_c} + \lambda_7\mathcal{L}_{con_p} + \lambda_8\mathcal{L}_{con_c_rd} + \lambda_9\mathcal{L}_{con_p_rd} \tag{8}$$

where $\lambda_1, \dots, \lambda_9$ act as trade-off parameters for different loss terms.

Although Eq. (8) contains nine weighted terms, these losses implement three coupled supervisory functions rather than nine independent modules. $L_{seg}$ provides task supervision for anatomical segmentation; $L_{recon}$ and $L_{rec_mk}$ preserve global appearance and task-relevant structure during reconstruction; and the six consistency terms constrain the re-encoding of original reconstructions, cross-image style exchanges, and randomly generated styles at the style, content, and prediction levels. The content and prediction consistency terms prevent a reassigned style from changing anatomical information, while the style consistency terms require the reconstructed image to express the assigned image-specific style. Together, these constraints operationalize OIOD by making every image-specific style code exchangeable while preserving anatomical content. Their roles should therefore be interpreted as a coupled objective for content-style separation rather than as nine unrelated mechanisms.

## 4. Experiment results and discussion

### 4.1. Datasets

In this study, we adopted two types of public challenge datasets for evaluating the proposed UniDDG framework: a multi-center fundus image dataset[1] [70-72] and a multi-center prostate T2-weighted MRI dataset [73-75].

*Multi-center fundus image dataset*: We evaluated our approach for Optic Cup (OC) and Optic Disc (OD) segmentation using a publicly available multi-center fundus image dataset. This dataset consists of images from four different public fundus image collections, each captured with different scanners at various sites, resulting in distinct domain discrepancies in visual appearance and image quality. Specifically:

a) Center 1 is from the Drishti-GS dataset [70], with 50 images for training and 51 for testing.
b) Center 2 is from the RIM-ONE-r3 dataset [71], containing 99 images for training and 60 for testing.
c) Center 3 and 4 are from the REFUGE challenge [72], with 320 training images and 80 testing images from the training and validation datasets, respectively.

A summary of the statistics for these multi-center fundus images is presented in Table 1.

Table 1 Overview of the fundus image statistics across the four centers used in our experiment.

| Task | Center No. | Dataset | Images (train / test) | Scanners |
|---|---|---|---|---|
| OC/OD segmentation | Center 1 | Drishti-GS [70] | 50 / 51 | Aravind eye hospital |
| | Center 2 | RIM-ONE-r3 [71] | 99 / 60 | Nidek AFC-210 |
| | Center 3 | REFUGE [72] (train) | 320 / 80 | Zeiss Visucam 500 |
| | Center 4 | REFUGE [72] (val) | 320 / 80 | Canon CR-2 |

In this study, it is important to note that the segmentation of the OD refers specifically to the segmentation of the OD region excluding the OC, as the OD anatomically encompasses the OC.

*Multi-center prostate T2-weighted MRI dataset*: We also evaluated our approach for prostate segmentation using a publicly available multi-center prostate T2-weighted MRI dataset. This dataset consists of five center-specific prostate T2-weighted MRI cohorts drawn from three public collections and acquired at different sites using distinct field strengths, spatial resolutions, and coil configurations, resulting in center-related variations in image appearance and quality. Specifically:

a) Center 1 and center 2 are from the NCI-ISBI13 dataset [73], and both centers contain 30 cases.
b) Center 3 is from the I2CVB dataset [74], including 19 cases.
c) Center 4 and center 5 are from the PROMISE12 dataset [75], with center 4 containing 12 cases and center 5 containing 13 cases.

For each of the five center datasets, we randomly divided the original site dataset into training and testing sets with a 4:1 ratio. A summary of the statistics for these multi-center prostate T2-weighted MRI cases is presented in Table 2. The prostate split was performed at the patient level before slice extraction, so that no patient contributes to both training and test sets.

Table 2 Overview of the prostate T2-weighted MRI case statistics across the five centers used in our experiment.

[1] https://github.com/emma-sjwang/Dofe

| Task | Center No. | Dataset | Cases (train / test) | Field strength (T) | Resolution (in/through plane) (mm) | Coil |
|---|---|---|---|---|---|---|
| Prostate segmentation | Center 1 | NCI-ISBI13 [73] | 24 / 6 | 3 | 0.6-0.625 / 3.6-4 | Surface |
| | Center 2 | NCI-ISBI13 [73] | 24 / 6 | 1.5 | 0.4 / 3.0 | Endorectal |
| | Center 3 | I2CVB [74] | 15 / 4 | 3 | 0.67-0.79 / 1.25 | - |
| | Center 4 | PROMISE12 [75] | 9 / 3 | 1.5 | 0.625 / 3.6 | Endorectal |
| | Center 5 | PROMISE12 [75] | 10 / 3 | 1.5 and 3 | 0.325-0.625 / 3-3.6 | - |

For preprocessing, we applied basic data augmentation (BDA) techniques (random flip, random rotate, random scale, random shift, random noise, random brightness, and random channel swap) to enhance the diversity of the training samples. During training, the fundus images were randomly cropped to a size of 256×256, and the prostate MRI slices were resized to 256×256. Note that fundus images are three-channel color images, while prostate MRI slices are single-channel grayscale images. To adapt the prostate MRI slices to the framework of this study, we replicated the single channel to convert them into three-channel images. Note that all background slices of prostate MRI datasets have been removed.

We assess the effectiveness of our proposed method from two perspectives: *first, the generalization from multiple source centers to a single target center, and second, the performance within a single center*. To evaluate the generalizability of the UniDDG framework across multiple source centers to a single target center, we employed a leave-one-center-out cross-validation strategy, where one center is held out for testing while the remaining centers are used for training. It is important to highlight that only the training image/slice sets from the multi-source centers are used for network training, while the testing image/slice sets from the target center are employed to assess the network's performance. Furthermore, we also evaluate the generalization ability of our method within a single center, where both the training and testing datasets are derived from the same center.

The evaluation includes nine center-specific datasets: four fundus photography datasets and five prostate T2-weighted MRI datasets. These datasets span two imaging modalities and include differences in acquisition sites, scanners, magnetic field strengths, spatial resolutions, and imaging protocols. The evaluation therefore examines center-related variation under both multi-source-to-unseen-center and within-center settings. This evidence supports the applicability of UniDDG to the two evaluated segmentation tasks, while validation on larger cohorts and additional imaging modalities is required to assess its applicability to broader forms of clinical heterogeneity.

### 4.2. Implementation details

Training and inference were carried out on a single NVIDIA GeForce RTX 2080 Ti GPU. It is worth noting that the Style Encoder and Decoder are used only during training. For inference, these modules are discarded, meaning the model's inference speed and memory usage are identical to the standard U-Net backbone. The content encoder $Econ$ was constructed using a U-Net [1] as the core architecture, with progressively increasing channel sizes of 16, 32, 64, 128, and 256 across five resolution scales. The number of channels for the content representations was fixed at $R$ = 8. The segmenter $S$ consists of two convolutional blocks: the first block includes a 3×3 convolution followed by batch normalization (BN) and LeakyReLU (with a slope of 0.2), while the second block features a 1×1 convolution followed by a Softmax layer to output a segmentation probability map. The style encoder $Esty$ contains convolutional layers with down-sampling steps to reduce spatial resolution, and its final output is passed through two fully connected layers to produce the mean and variance for the latent style code, with the size set to $Z$ = 48.

As shown in Table 3, we further analyzed the architecture-level computational cost of UniDDG using a $256 \times 256$input. FLOPs were calculated for convolutional and fully connected operations by counting one multiplication and one addition as two operations. During training, UniDDG contains 3.44 M trainable parameters and its forward computation requires approximately 26.76 GFLOPs per image. During inference, the style encoder and decoder are removed, reducing the deployed model to the content encoder and segmentation head, with 2.20 M parameters and 8.54 GFLOPs per image. The corresponding FP32 parameter storage decreases from 13.76 MB during training to 8.79 MB during inference. We further report the measured runtime of UniDDG on the single NVIDIA GeForce RTX 2080 Ti GPU used in this study: total training time of approximately 5.45 hours for the fixed 500-epoch schedule, peak GPU memory of 4.62 GB during training, and convergence around epoch 82, after which the source-validation performance plateaued. These figures are hardware- and implementation-specific and are reported for transparency; they are not used for cross-method runtime benchmarks.

Table 3 Architecture-level computational complexity of UniDDG.

| Phase | Retained modules | Parameters | Forward FLOPs | FP32 parameter storage |
|---|---|---|---|---|
| Training | $E_{con}$, $E_{sty}$, $S$, and $D$ | 3.44 M | 26.76 G | 13.76 MB |
| Inference | $E_{con}$ and $S$ | 2.20 M | 8.54 G | 8.79 MB |

The number of model parameters and the per-sample computation graph are independent of $K$ because UniDDG does not instantiate domain-specific encoders, decoders, or segmentation networks. For a fixed batch size, adding source centers changes only the composition and potentially the number of training samples, rather than the model architecture or computational cost of an individual optimization step.

The loss function weights were determined using a prioritized step-wise tuning strategy. We first optimized the primary task weights ($\lambda_1$, $\lambda_2$, $\lambda_3$, $\lambda_5$) to ensure stable segmentation and reconstruction, and then incrementally introduced the other regularization weights to enforce disentanglement and segmentation. The final values were set as follows: $\lambda_1$=5.0, $\lambda_2$=5.0, $\lambda_3$ =10.0, $\lambda_4$=1.0, $\lambda_5$=5.0, $\lambda_6$=1.0, $\lambda_7$=1.0, $\lambda_8$=1.0, $\lambda_9$=1.0. The coefficient selection follows the functional hierarchy described in Section 3.7. $L_{seg}$ uses Dice loss, whereas the reconstruction and consistency terms use L1 distance; thus, the nine supervision terms are implemented using only two loss forms. During prioritized step-wise tuning, $\lambda_1$, $\lambda_2$, $\lambda_3$, and $\lambda_5$ were determined first, and the remaining five regularization weights were then introduced at 1.0. The resulting coefficient vector was kept fixed across the fundus and prostate experiments in both multi-source and single-source settings, without setting-specific retuning. Therefore, all results in Tables 4–11 were obtained using one shared loss configuration. The loss terms are evaluated only during training and introduce no inference-time hyperparameters. The models were optimized using the RMSprop optimizer, with an initial learning rate of $10^{-4}$. For each mini-batch, the number of images/slices was set as 8. The number of epochs was set as 500. Segmentation performance was evaluated using two metrics: the Dice coefficient (Dice) and the Average Symmetric Surface Distance (ASSD).

The prioritized step-wise strategy was a function-guided sequential calibration rather than an exhaustive joint grid or Bayesian search over the nine-dimensional coefficient space. Specifically, the four coefficients directly associated with segmentation and reconstruction ($\lambda_1$, $\lambda_2$, $\lambda_3$, and $\lambda_5$) were determined first, whereas $\lambda_4$ and $\lambda_6$–$\lambda_9$ were subsequently introduced at the fixed value of 1.0. Thus, only four coefficients were actively adjusted, rather than all nine coefficients being treated as independent search dimensions. Once determined, the complete coefficient vector was fixed and applied unchanged to all fundus and prostate experiments, including both multi-source and single-center settings. No dataset-specific, center-specific, or evaluation-setting-specific retuning was performed. Consequently, all experiments in Tables 4–11 can be reproduced using the single coefficient configuration reported above. For reproducibility, the tuning procedure is specified as follows: (1) search ranges — the four actively tuned coefficients λ1, λ2, λ3, and λ5 were explored over {1.0, 2.5, 5.0, 10.0} in the prioritized order described above, while λ4 and λ6–λ9 were introduced at the fixed value of 1.0; (2) validation protocol — coefficient selection used a small held-out subset of the source training data only, and the target center was excluded, consistent with the leave-one-center-out protocol; (3) stopping criteria — a fixed schedule of 500 epochs without early stopping; and (4) search budget —the tuning was performed manually in a prioritized, step-wise manner: each candidate configuration varied a single coefficient while the others were held fixed, was evaluated on the held-out source subset, and was retained only when it improved the validation Dice. In total, 15 configurations were examined before the final coefficient vector was fixed.

For each leave-one-center-out fold, the held-out center was excluded from hyperparameter selection, model selection, and stopping decisions. The loss weights, learning rate, training duration, and remaining implementation settings were fixed before evaluation and were shared across all folds. Data and performance measurements from the held-out center were used only for final testing. For every leave-one-center-out fold, the validation subset used for coefficient selection was drawn exclusively from the source centers of that fold; the target center never contributed to coefficient selection, early stopping or model selection. Each setting was trained once under the reported configuration; we do not claim seed-level stability.

Statistical comparisons were performed between UniDDG and the strongest comparison method in each experimental setting using a two-sided paired Wilcoxon signed-rank test on the per-case Dice scores obtained by the two methods for the same evaluation cases. This paired design evaluates whether the observed differences are consistently signed across cases rather than relying only on the aggregated mean scores reported in Tables 4–11. Statistical significance was assessed at $\alpha = 0.05$. The resulting performance differences and $p$ values are summarized in Table 14. The test was run over N = 271 held-out test images in the two fundus settings (51 + 60 + 80 + 80) and N = 22 held-out cases in the two prostate settings (6 + 6 + 4 + 3 + 3), where the per-case Dice of a prostate case is computed on the 3D volume obtained by stacking its predicted slices against the corresponding ground-truth volume.

## 4.3. Fundus image segmentation

### *4.3.1. Generalization from multiple source centers to a single target center*

(i) Comparison with state-of-the-art DG methods: For evaluating the generalization performance of the proposed UniDDG framework from multiple source centers to a single target center, a leave-one-center-out cross-validation strategy is employed. Initially, we treated all available training centers as a unified dataset, disregarding domain shifts within the training set, and trained a U-Net [1] using standard Dice loss. This model was then directly applied to an unseen center, serving as the experiment's lower bound. Instead, for each center, we trained and tested only in their data set, ensuring that no invisible centers were involved. This setting determines the upper bound of DG. In addition, to provide a more refined benchmark, we have excluded basic data enhancement (BDA) operations in both Settings. This led to the establishment of an upper bound w/o BDA and a lower bound w/o BDA as an additional reference, resulting in a clearer understanding of the impact of domain shifting and the benefits of data enhancement. To evaluate DG performance, we benchmarked our proposed UniDDG against eight cutting-edge approaches: Cutmix [56], Mixup [76], BigAug [24], Manifold Mixup [55], Dofe [12], CDDSA [13], TriD [61] and DFQ [62]. These methods represent various strategies for enhancing domain generalization, including CutMix[56] and Mixup[76], which generate enhanced samples by mixing images, BigAug[24], which simulates domain migration by stacking different transformations, and interpolating in feature space to create a domain invariant representation of Manifold Mixup[55]. Dofe[12] uses domain-oriented feature embedding to perform robust segmentation across multiple domains, and CDDSA[13] uses contrast domain de-entanglement and Style Augmentation to enhance generalization. TriD [61] utilizes domain randomization to synthesize various imaging conditions to extract domain invariant features, although its simulated shifts may not fully reflect the complexity of the real world. At the same time, DFQ [62] decouple semantic and spatial feature extraction to mitigate overfitting, but this strategy introduces additional computational overhead and presents challenges in seamlessly integrating the decouple features. For the multi-source setting in Table 4 and Table 5, we further compared UniDDG with two recent methods: MixStyleFlow [64], which employs normalizing flows for feature-style modeling, and Aegis [65], which mitigates cross-domain feature misalignment. DAPSAM [63] is a SAM-based method designed for single-source generalization and is therefore evaluated under the corresponding single-center protocols in Table 6, Table 7, Table 10, and Table 11. Note that all DG methods, including UniDDG, are executed using BDA.

Table 4 Evaluation of Dice Scores (%) ↑ using diverse DG approaches on fundus image dataset from multiple source centers to a single target center

| Method-Year | Center1 | | Center2 | | Center3 | | Center4 | | Average | | |
|---|---|---|---|---|---|---|---|---|---|---|---|
| | OC | OD | OC | OD | OC | OD | OC | OD | OC | OD | All |
| Lower bound w/o BDA | 76.55 | 74.06 | 67.91 | 64.35 | 81.10 | 84.28 | 77.89 | 69.29 | 75.86 | 73.00 | 74.43 |
| Lower bound | 80.36 | 79.94 | 74.68 | 71.41 | 84.97 | 84.32 | 85.09 | 85.68 | 81.28 | 80.34 | 80.81 |
| Upper bound w/o BDA | 85.89 | 83.85 | 80.73 | 84.56 | 87.59 | 90.45 | 89.54 | 90.81 | 85.94 | 87.42 | 86.68 |
| Upper bound | 85.59 | 84.11 | 80.23 | 83.55 | 88.40 | 90.63 | 90.00 | 91.50 | 86.06 | 87.45 | 86.75 |
| Cutmix-2019 [56] | 85.22 | 78.65 | 76.94 | 77.36 | 85.81 | 85.21 | 84.30 | **88.34** | 83.07 | 82.39 | 82.73 |
| Mixup-2017 [76] | 80.44 | 74.02 | 71.90 | 70.95 | 86.02 | 80.07 | 83.56 | 85.63 | 80.48 | 77.67 | 79.07 |
| BigAug-2020 [24] | 83.47 | 77.73 | 78.73 | 77.02 | 85.75 | 83.11 | 84.97 | 86.82 | 83.23 | 81.17 | 82.20 |
| Manifold Mixup-2019 [55] | 76.87 | 70.43 | 78.67 | 75.31 | 81.04 | 80.81 | 85.06 | 87.58 | 80.41 | 78.53 | 79.47 |
| Dofe-2020 [12] | 85.13 | 78.94 | 81.00 | **81.10** | **87.39** | 85.94 | 84.96 | 86.80 | 84.62 | 83.20 | 83.91 |
| Dofe w/ wrong domain label | 83.91 | 81.74 | 78.99 | 78.96 | 83.97 | 85.80 | 84.78 | 87.91 | 82.91 | 83.60 | 83.26 |
| CDDSA-2023 [13] | 83.46 | 82.52 | 78.74 | 74.44 | 85.77 | 86.28 | 86.28 | 85.26 | 83.56 | 82.13 | 82.84 |
| CDDSA w/ wrong domain label | 79.33 | 75.84 | 76.27 | 73.28 | 85.00 | 85.94 | 84.91 | 85.87 | 81.38 | 80.23 | 80.80 |
| TriD-2023 [61] | 83.27 | 79.74 | 80.44 | 77.81 | 83.41 | 82.09 | 85.44 | 86.22 | 83.14 | 81.46 | 82.30 |
| DFQ-2024 [62] | **86.62** | 82.31 | **83.04** | 78.32 | 85.31 | 85.48 | 85.63 | 88.17 | **85.15** | 83.57 | 84.36 |
| MixStyleFlow-2025 [64] | 78.25 | 74.15 | 77.52 | 72.41 | 83.25 | 84.72 | 85.52 | 83.17 | 81.14 | 78.61 | 79.87 |
| Aegis-2025 [65] | 84.93 | 82.2 | 79.01 | 76.71 | 86.35 | 87.17 | 86.85 | 88.54 | 84.28 | 83.65 | 83.97 |

| | | | | | | | | | | |
|---|---|---|---|---|---|---|---|---|---|---|
| UniDDG w/o EMA | 83.61 | 80.25 | 78.48 | 75.70 | 83.56 | 81.88 | 85.94 | 87.37 | 82.90 | 81.30 | 82.10 |
| UniDDG w/o SA | 84.14 | 81.71 | 79.62 | 76.47 | 86.85 | 84.51 | **86.95** | 88.11 | 84.39 | 82.70 | 83.55 |
| **UniDDG (Ours)** | 85.54 | **82.69** | 80.07 | 76.93 | 87.16 | **87.92** | 86.93 | 88.22 | 84.93 | **83.94** | **84.43** |

Table 4 and Table 5 present the quantitative evaluation results of OC and OD segmentation in terms of Dice and ASSD, respectively, for the various DG methods. The inclusion of BDA played a pivotal role in these experiments, particularly when comparing the upper and lower bounds with and without BDA. The results clearly highlight the importance of BDA in improving model generalization. For the lower bound, the Dice score increased from 74.43% (without BDA) to 80.81% (with BDA), and the ASSD improved from 8.28 pixels to 4.52 pixels, indicating that enhancing the visual diversity of training data significantly aids in model generalization. However, when considering the upper bound, the improvement from 86.68% (without BDA) to 86.75% (with BDA) was marginal, suggesting that in settings where domain shifts are less severe, the contribution of BDA to improving generalization is limited. This finding emphasizes that while BDA provides essential benefits in more challenging domain generalization tasks, its impact is less pronounced when the domain shifts are relatively small. Therefore, these results validate the importance of BDA as a fundamental strategy for enhancing model robustness, while also pointing out that its effects are most substantial when domain shifts are more significant.

Table 5 Evaluation of ASSD (pixel) ↓ using diverse DG approaches on fundus image dataset from multiple source centers to a single target center

| Method-Year | Center1 | | Center2 | | Center3 | | Center4 | | Average | | |
|---|---|---|---|---|---|---|---|---|---|---|---|
| | OC | OD | OC | OD | OC | OD | OC | OD | OC | OD | All |
| Lower bound w/o BDA | 7.53 | 4.60 | 13.12 | 10.36 | 4.38 | 3.82 | 5.53 | 16.88 | 7.64 | 8.92 | 8.28 |
| Lower bound | 6.49 | 3.80 | 6.00 | 7.04 | 3.72 | 3.59 | 2.65 | 2.86 | 4.72 | 4.32 | 4.52 |
| Upper bound w/o BDA | 4.64 | 3.23 | 4.09 | 3.81 | 2.89 | 2.17 | 1.83 | 1.68 | 3.36 | 2.72 | 3.04 |
| Upper bound | 4.77 | 3.07 | 4.24 | 3.55 | 2.73 | 2.11 | 1.75 | 1.58 | 3.37 | 2.58 | 2.97 |
| Cutmix-2019 [56] | 5.13 | 3.84 | 6.06 | 4.70 | 3.19 | 3.12 | 2.72 | 2.14 | 4.27 | 3.45 | 3.86 |
| Mixup-2017 [76] | 6.35 | 4.54 | 12.64 | 10.90 | 3.59 | 4.12 | 3.13 | 2.54 | 6.42 | 5.53 | 5.97 |
| BigAug-2020 [24] | 5.51 | 3.91 | 4.48 | 4.51 | 3.38 | 3.52 | 2.66 | 2.34 | 4.01 | 3.57 | 3.79 |
| Manifold Mixup-2019 [55] | 7.57 | 4.96 | 4.82 | 4.68 | 4.21 | 4.20 | 2.63 | **2.13** | 4.81 | 3.99 | 4.40 |
| Dofe-2020 [12] | 4.83 | 4.52 | **4.03** | **3.68** | **2.91** | 2.90 | 2.46 | 2.29 | 3.56 | **3.35** | 3.45 |
| Dofe w/ wrong domain label | 5.22 | 4.22 | 4.62 | 4.45 | 3.56 | 2.94 | 3.60 | 2.17 | 4.25 | 3.44 | 3.85 |
| CDDSA-2023 [13] | 5.49 | **2.14** | 4.74 | 5.89 | 3.18 | 2.78 | 2.27 | 2.75 | 3.92 | 3.39 | 3.65 |
| CDDSA w/ wrong domain label | 6.78 | 3.50 | 5.37 | 5.64 | 3.67 | 3.37 | 2.62 | 3.06 | 4.61 | 3.89 | 4.25 |
| TriD-2023 [61] | 3.86 | 5.91 | 4.58 | 4.17 | 3.75 | 3.60 | 2.46 | 2.64 | 3.66 | 4.08 | 3.87 |
| **DFQ-2024** [62] | **3.14** | 4.42 | 4.64 | 3.77 | 3.06 | 3.24 | **2.12** | 2.54 | **3.24** | 3.49 | **3.37** |
| MixStyleFlow-2025 [64] | 3.83 | 6.89 | 5.72 | 5.21 | 3.10 | 3.77 | 2.63 | 2.62 | 3.82 | 4.62 | 4.22 |
| Aegis-2025 [65] | 3.24 | 4.33 | 4.92 | 4.40 | 3.03 | 3.19 | 2.42 | 2.63 | 3.40 | 3.64 | 3.52 |
| UniDDG w/o EMA | 5.41 | 3.78 | 6.13 | 5.16 | 4.08 | 5.24 | 3.00 | 3.81 | 4.66 | 4.50 | 4.58 |
| UniDDG w/o SA | 5.48 | 3.69 | 4.54 | 5.02 | 3.09 | 3.69 | 2.38 | 2.27 | 3.87 | 3.67 | 3.77 |
| UniDDG (Ours) | 4.79 | 3.20 | 4.35 | 5.02 | 2.93 | **2.58** | 2.35 | 5.88 | 3.61 | 4.17 | 3.89 |

UniDDG consistently achieved the best performance, with an average Dice score of 84.43%, outperforming state-of-the-art DG methods such as CutMix (82.73%), Mixup (79.07%), BigAug (82.20%), Manifold Mixup (79.47%), Dofe (83.91%), CDDSA (82.84%), TriD (82.30%), DFQ (84.36%), and Aegis (83.97%), which also significantly outperform the lower bound (80.81%). The improvement over DFQ, despite its strong performance, highlights the effectiveness of UniDDG in capturing and leveraging domain-invariant features while maintaining robust generalization. Similarly, MixStyleFlow employs normalizing flows to model domain styles; however, it achieved a lower Dice score (79.87%), suggesting that explicit flow-based modeling may struggle with the extreme scanner variability found in fundus imaging. Aegis, while delivering a competitive Dice score (83.97%) by mitigating cross-domain feature misalignment, still falls slightly short of UniDDG (84.43%). This indicates that our "One Image as One Domain" strategy offers a more robust generalization mechanism than explicitly aligning misaligned features in high-dimensional space. UniDDG achieves the best average Dice (84.43%) and an ASSD of 3.89 pixels, which lies within the range of the strongest baselines (DFQ 3.37 and Aegis 3.20 pixels); we therefore claim competitive rather than uniformly superior boundary accuracy. Although DFQ achieved a slightly lower ASSD (3.37 pixels), its Dice score was noticeably inferior. This observation highlights the trade-off between boundary precision and overall segmentation quality. Dice score, being a more holistic metric for segmentation quality, holds greater importance in practical applications, especially when global segmentation accuracy is prioritized.

Comparing UniDDG with other DG methods demonstrates its advantages. Techniques like Mixup and CutMix, which are effective in natural image domains, struggle to preserve the structural and spatial complexities of medical images. Methods such as BigAug and Manifold Mixup lack sufficient adaptability to domain-specific challenges in medical imaging. While Dofe's domain-oriented feature embedding improves boundary precision, it falls short in achieving competitive Dice scores, reflecting its limitations in overall segmentation quality. CDDSA, which employs contrastive domain disentanglement and style augmentation, also shows limitations, with lower Dice scores and higher ASSD compared to UniDDG. TriD utilizes domain randomization to generate diverse imaging conditions, aiming to extract domain-invariant features. However, its reliance on simulated shifts may not fully capture real-world complexities, leading to less robust generalization. DFQ decouples semantic and spatial feature extraction to alleviate overfitting, but this approach increases computational complexity and struggles to integrate features seamlessly. UniDDG's strong performance across both metrics highlights its robustness in multi-source-to-single-target scenarios, even under significant domain shifts. The framework effectively balances segmentation accuracy (Dice score) with boundary precision (ASSD), surpassing competing methods on Dice, while remaining competitive on ASSD. Visual comparisons (Figure 6) further confirm that UniDDG achieves the largest overlap with the ground truth in the segmented regions, with fewer discrepancies in segmentation compared to other DG methods.

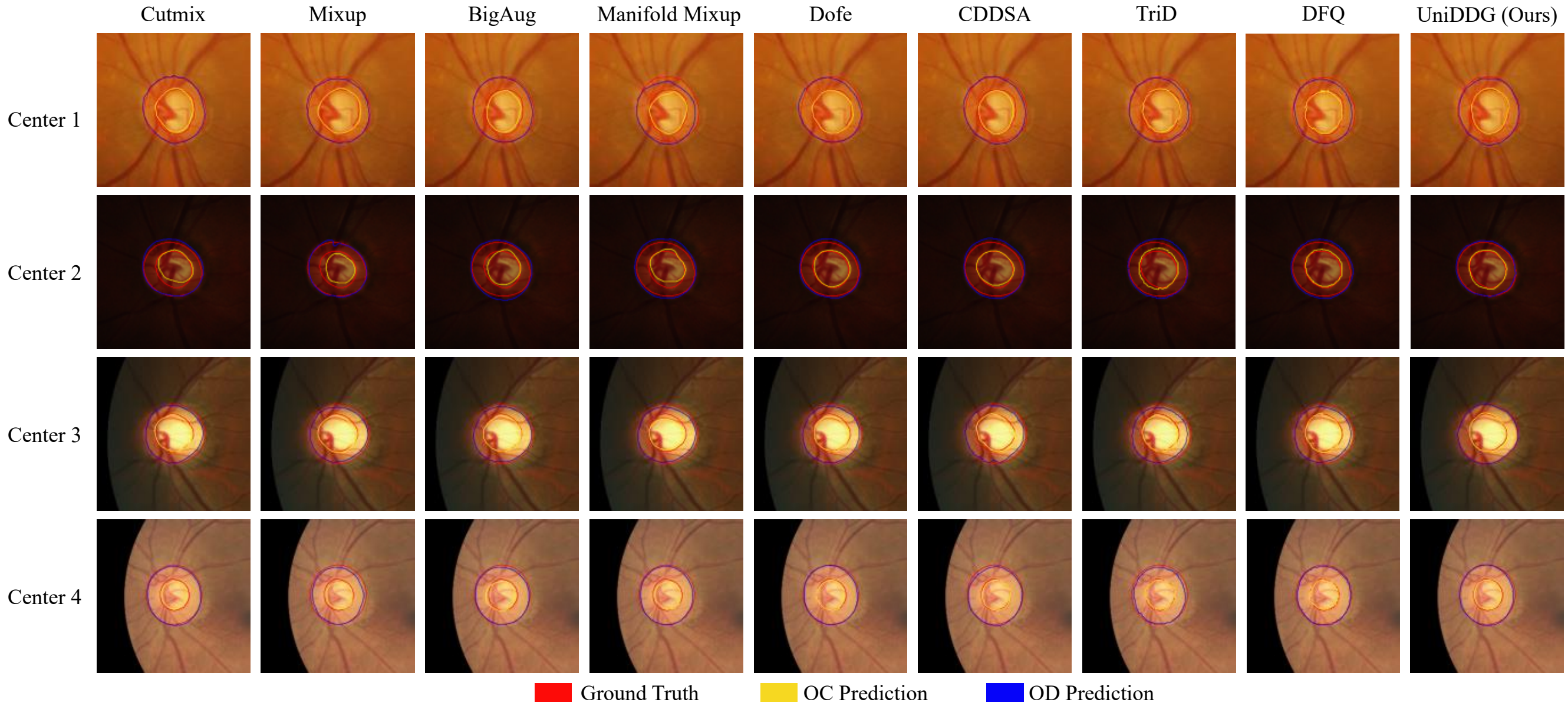

Figure 6 Visualization of segmentation results using diverse DG approaches on fundus image dataset from multiple source centers to a single target center

(ii) Ablation studies: This section investigates the contributions of the EMA and SA modules to the performance of the UniDDG framework through ablation studies. To achieve this, we compare three model variants: UniDDG w/o EMA, UniDDG w/o SA, and the full UniDDG model. These variants were created by individually removing the EMA and SA modules, and their performance was evaluated quantitatively, as shown in Table 4 and Table 5.

The UniDDG w/o EMA variant, which excludes the expansion mask attention (EMA) module, shows a significant drop in performance. Specifically, the Dice score decreased from 84.43% in the full UniDDG model to 82.10%, while the Average Symmetric Surface Distance (ASSD) increased from 3.89 pixels to 4.58 pixels. These results underscore the importance of the EMA module in improving segmentation accuracy and preserving boundary details. To better understand this, Figure 7 provides visualizations of reconstructed images from a single batch during training for both the full UniDDG model and the UniDDG w/o EMA configuration. In the full UniDDG setting, the reconstructed images $x'_i$, $x'_{i\to j}$, and $x'_{i\to rd}$ successfully retain the complete structures of the optic disc, optic cup, and their surrounding regions, ensuring that critical anatomical features are preserved for effective segmentation. *Meanwhile, as can be seen in Figure 7, our model effectively disentangles the content and style of images, allowing for style exchange among images within the same batch. By comparing* $x_i$ *and* $x'_{i\to j}$*, the visual differences after style swapping are clearly observable.* In this controlled comparison, $x'_{i\to j}$ is generated by keeping the content representation $c_i$ unchanged and replacing only its associated style code. The preservation of the optic disc and optic cup geometry, together with the observable appearance changes, provides a functional visualization of anatomical content preservation and style variation. The reconstructions generated with random style codes, $x'_{i\to rd}$, further show that the content representation remains stable under styles outside the observed batch combinations. This exchange promotes the fusion of multiple styles with each structure, thereby enhancing the model's generalization ability.

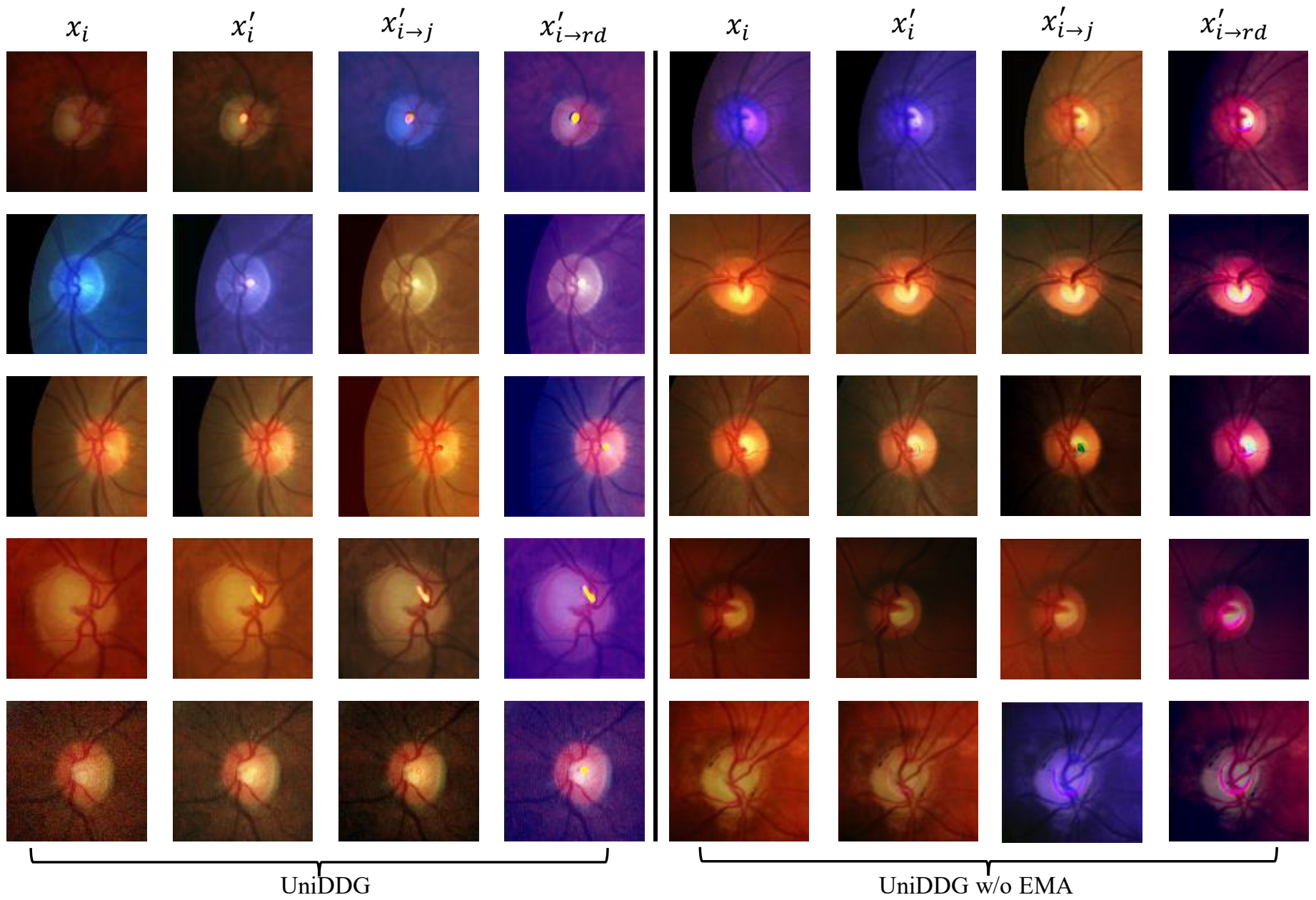

Figure 7 Visualization of reconstructed images of one batch during training process (note this is the last batch in an epoch, so batch size is not the 8 mentioned earlier). $x_i$, $x'_i$, $x'_{i\to j}$, and $x'_{i\to rd}$ represent original image, reconstruction of original image, reconstructed image of exchanging styles within a batch, and reconstructed image with a random style

In contrast, the UniDDG w/o EMA configuration produces reconstructed images that exhibit prominent artificial contour lines around the boundaries of the optic disc and optic cup. These contour lines are introduced into the training process through the content encoder and segmentation head, significantly affecting the learning dynamics of the model. This comparison provides clear feature-level visual evidence for the contribution of the EMA module to edge processing. Without the region-specific attention provided by EMA, the content decoder would embed boundary information of the optic cup and optic disc into the content codes during training to maximize segmentation accuracy on the training data, so that images reconstructed by combining these content codes with any style codes can be easily segmented by the segmentation network, resulting in clear boundary information in all reconstructed images (as $x'_i$, $x'_{i\to j}$, and $x'_{i\to rd}$ in Figure 7). This phenomenon causes the segmentation network to rely excessively on the explicit boundary information encoded in the content codes rather than learning to extract the optic cup and optic disc structures from raw images without clear boundaries, thus leading to a significant performance drop on the test dataset, as the test data contain no explicit boundary information. The EMA technique enforces that the reconstructed images share identical structure with the raw images within the EMA-attended regions during training, thereby eliminating the artificial clear boundaries caused by model shortcut learning in the reconstructed images and further removing the boundary information embedded in the content codes, which enables the model to achieve favorable performance on the test dataset. Thus, removal of EMA resulted in a significant decline in quantitative measures and qualitative reconstruction results, underscoring its critical role in segmentation accuracy and generalization.

For the UniDDG w/o SA variant that does not include the Style Augmentation (SA) module, the results show a slight improvement in boundary accuracy, with ASSD reduced from 3.89 pixels to 3.77 pixels. However, the Dice score also decreased slightly, from 84.43% to 83.55%, indicating a slight decline in overall segmentation accuracy. The SA module introduces different style information in the training process (as shown by $x'_{i\to rd}$ in Figure 7), which enhances the robustness of the model to styles in different domains. While removing the SA module reduces the variability of boundary accuracy, it slightly hinders the model's ability to adapt to cross-domain style changes. This adaptability is essential for maintaining robust performance under a variety of conditions, even if it comes at the cost of slightly reduced edge detection accuracy.

In summary, ablation studies have shown that EMA and SA modules contribute uniquely to the performance of UniDDG models. The EMA module is essential for capturing fine boundary details and ensuring accurate segmentation, while the SA module enhances the model's adaptability to cross-domain style changes. The removal of EMA results in a significant decrease in performance, while the removal of SA slightly reduces overall accuracy while improving boundary accuracy. These modules complement each other in the complete UniDDG model, enabling it to achieve high segmentation accuracy and robustness to domain shift.

*4.3.2. Generalization within a single center*

(i) Comparison with state-of-the-art DG methods: In order to evaluate the generalization performance of the proposed UniDDG framework within a single center, both the training and testing datasets are from the same center. We use the standard Dice loss as the lower bound to train U-Net [1], which serves as the upper bound for generalization evaluation from multiple source centers to a single target center, as shown in Table 4 and Table 5. Consequently, the results of the upper bound and upper bound w/o BDA from these tables are directly used as the lower bound and lower bound w/o BDA for the experiments in this section. In comparison with state-of-the-art domain generalization (DG) methods, we focus on the methods most suitable for this scenario: UniDDG, Cutmix, Mixup, BigAug, and Manifold Mixup. Dofe and CDDSA, which were designed specifically for multi-source center scenarios, are not appropriate for this single-center experiment and, therefore, is excluded from this comparison.

Table 6 and Table 7 present the quantitative evaluation results for OC and OD segmentation in terms of Dice score and ASSD, respectively, comparing various domain generalization (DG) methods within a single-center scenario. The results demonstrate that the proposed UniDDG method outperforms the lower bound, as well as several state-of-the-art DG methods, confirming its superior generalization capability.

In the Dice score evaluation (Table 6), UniDDG achieved an average score of 88.91%, which is a significant improvement over the lower bound (86.75%). This indicates that UniDDG effectively improves generalization and achieves high-quality segmentation. Compared to other methods, UniDDG consistently performs better. For instance, Cutmix and Mixup achieved average Dice scores of 87.38% and 86.89%, respectively, while BigAug and Manifold Mixup achieved scores of 87.17% and 85.57%, respectively. We also compared UniDDG with DAPSAM, a recent method leveraging prompt learning. While DAPSAM achieves a high Dice score of 88.80%, UniDDG still maintains a slight edge (88.91%). This suggests that our style-content disentanglement remains highly effective even within a single center, achieving performance comparable to or better than complex foundation-model-based prompting strategies. These results highlight UniDDG's ability to achieve more accurate segmentation than other methods that rely on data augmentation techniques.

Table 6 Evaluation of Dice Scores (%) ↑ using diverse DG approaches on fundus image dataset within a single center

| Method-Year | Center1 | | Center2 | | Center3 | | Center4 | | Average | | |
|---|---|---|---|---|---|---|---|---|---|---|---|
| | OC | OD | OC | OD | OC | OD | OC | OD | OC | OD | All |
| Lower bound w/o BDA | 85.89 | 83.85 | 80.73 | 84.56 | 87.59 | 90.45 | 89.54 | 90.81 | 85.94 | 87.42 | 86.68 |
| Lower bound | 85.59 | 84.11 | 80.23 | 83.55 | 88.40 | 90.63 | 90.00 | 91.50 | 86.06 | 87.45 | 86.75 |
| Cutmix-2019 [56] | 85.36 | 85.02 | 83.40 | 85.41 | 88.28 | 90.65 | 89.62 | 91.29 | 86.67 | 88.09 | 87.38 |
| Mixup-2017 [76] | 85.88 | 85.14 | 81.74 | 84.78 | 87.93 | 90.27 | 88.66 | 90.73 | 86.05 | 87.73 | 86.89 |
| BigAug-2020 [24] | 86.72 | 83.58 | 82.49 | 85.73 | 87.96 | 90.63 | 89.29 | 90.92 | 86.62 | 87.72 | 87.17 |
| Manifold Mixup-2019 [55] | 86.21 | 84.50 | 74.87 | 81.31 | 88.06 | 90.16 | 88.93 | 90.50 | 84.52 | 86.62 | 85.57 |
| DAPSAM-2024 [63] | 88.57 | **87.96** | 84.08 | 88.17 | 88.70 | 90.84 | 90.25 | 91.82 | 87.90 | **89.69** | 88.80 |
| UniDDG w/o EMA | 85.59 | 84.74 | 84.39 | 86.72 | 88.89 | 90.92 | 90.17 | 91.96 | 87.26 | 88.59 | 87.92 |
| UniDDG w/o SA | 88.64 | 87.20 | **84.89** | **87.48** | 88.50 | 89.56 | 90.34 | 92.08 | 88.09 | 89.08 | 88.59 |
| **UniDDG (Ours)** | **89.01** | 87.27 | **84.89** | 87.42 | **88.95** | **91.03** | **90.66** | **92.03** | **88.38** | 89.44 | **88.91** |

For the ASSD evaluation (Table 7), UniDDG achieved an average of 2.52 pixels, outperforming the lower bound (2.97 pixels) and demonstrating better boundary delineation and precision. In comparison, Cutmix, Mixup, BigAug, and Manifold Mixup achieved average ASSD values of 2.76, 2.83, 2.76, and 3.17, respectively, indicating that UniDDG provides better boundary accuracy. These results further confirm that UniDDG excels in both segmentation accuracy (Dice score) and boundary precision (ASSD), surpassing the performance of other methods.

While data augmentation techniques such as Cutmix, Mixup, BigAug, and Manifold Mixup have been widely used to improve model generalization, their performance in this single-center scenario is relatively limited. This is likely due to the fact that domain shifts between the training and testing datasets are relatively small in this setting. Although domain shifts do exist (as indicated by the lower bound results), they are not as severe as in multi-center scenarios, and the performance improvement from these augmentation techniques is constrained. In contrast, UniDDG's domain generalization approach, which leverages advanced techniques to address domain discrepancies, consistently outperforms these state-of-the-art augmentation-based methods.

Table 7 Evaluation of ASSD (pixel) ↓ using diverse DG approaches on fundus image dataset within a single center

| Method-Year | Center1 | | Center2 | | Center3 | | Center4 | | Average | | |
|---|---|---|---|---|---|---|---|---|---|---|---|
| | OC | OD | OC | OD | OC | OD | OC | OD | OC | OD | All |
| Lower bound w/o BDA | 4.64 | 3.23 | 4.09 | 3.81 | 2.89 | 2.17 | 1.83 | 1.68 | 3.36 | 2.72 | 3.04 |
| Lower bound | 4.77 | 3.07 | 4.24 | 3.55 | 2.73 | 2.11 | 1.75 | 1.58 | 3.37 | 2.58 | 2.97 |
| Cutmix-2019 [56] | 4.76 | 2.81 | 3.55 | 2.86 | 2.71 | 2.05 | 1.76 | 1.56 | 3.19 | 2.32 | 2.76 |
| Mixup-2017 [76] | 4.63 | 2.79 | 4.05 | 2.79 | 2.80 | 2.06 | 1.89 | 1.64 | 3.34 | 2.32 | 2.83 |
| BigAug-2020 [24] | 4.43 | 2.86 | 3.79 | 2.75 | 2.77 | 2.02 | 1.84 | 1.60 | 3.21 | 2.30 | 2.76 |
| Manifold Mixup-2019 [55] | 4.51 | 2.84 | 6.38 | 3.29 | 2.72 | 2.06 | 1.85 | 1.71 | 3.86 | 2.47 | 3.17 |
| DAPSAM-2024 [63] | 3.87 | 2.54 | 3.84 | **2.20** | 2.67 | **1.95** | 1.69 | 1.71 | 3.01 | **2.10** | 2.56 |
| UniDDG w/o EMA | 4.69 | 2.94 | 3.46 | 2.42 | 2.86 | 2.05 | 1.70 | 1.48 | 3.18 | 2.22 | 2.70 |
| UniDDG w/o SA | 3.71 | **2.35** | **3.32** | 2.38 | 2.73 | 3.19 | 1.70 | **1.45** | 2.86 | 2.34 | 2.60 |
| **UniDDG (Ours)** | **3.57** | 2.52 | 3.46 | 2.85 | **2.57** | 2.00 | **1.64** | 1.57 | **2.81** | 2.23 | **2.52** |

(ii) Ablation studies: We conducted ablation experiments, where we systematically removed the EMA and SA modules from the UniDDG framework. The quantitative evaluation results are summarized in Table 6 and Table 7. The results show that removing either module leads to a decrease in performance. Specifically, UniDDG w/o EMA exhibited a noticeable performance drop, with the Dice score decreasing from 88.91% to 87.92%, and the ASSD increasing from 2.52 pixels to 2.70 pixels. Similarly, UniDDG w/o SA also resulted in a slight decline in Dice score from 88.91% to 88.59%, and a slight decline in ASSD, increasing it from 2.52 pixels to 2.60 pixels. These results highlight the importance and effectiveness of both modules: both EMA and SA play a crucial role in stabilizing training, enhancing segmentation accuracy, and improving the model's ability to adapt to domain shifts.

### 4.4. Prostate MRI segmentation

*4.4.1. Generalization from multiple source centers to a single target center*

(i) Comparison with state-of-the-art DG methods: To further validate the effectiveness of the UniDDG framework for medical image segmentation, we conducted experiments on a prostate MRI dataset, employing the same evaluation protocol as that used for the fundus segmentation dataset (see section 4.3.1). The results from the prostate MRI dataset reaffirm the robustness and adaptability of UniDDG in addressing domain shifts in medical imaging tasks.

Table 8 and Table 9 present the comparison results for the prostate MRI segmentation. In the lower bound setup, domain shifts between the centers are significant, leading to a relatively larger gap in performance. On the other hand, in the upper bound setup, where training and testing data come from the same center, the domain shift is relatively smaller, resulting in higher performance. The results show that when basic data augmentation (BDA) is applied, the Dice scores and ASSD values improve significantly in the lower bound scenario, where domain shifts are more severe. However, the improvements are relatively small in the upper bound scenario, where the domain shifts are weaker. This trend mirrors the findings in fundus image segmentation, where BDA has a significant impact in cases with substantial domain shifts and only a marginal effect when the domain shift is small.

Table 8 Evaluation of Dice Scores (%) ↑ using diverse DG approaches on prostate MRI dataset from multiple source centers to a single target center

| Method-Year | Center1 | Center2 | Center3 | Center4 | Center5 | Average |
|---|---|---|---|---|---|---|
| Lower bound w/o BDA | 84.43 | 74.19 | 80.41 | 74.52 | 83.40 | 79.39 |
| Lower bound | 85.17 | 83.36 | 83.78 | 78.01 | 84.28 | 82.92 |
| Upper bound w/o BDA | 88.08 | 86.02 | 86.59 | 83.00 | 77.47 | 84.23 |
| Upper bound | 88.00 | 88.53 | 89.61 | 81.74 | 78.45 | 85.27 |
| Cutmix-2019 [56] | 85.72 | 80.88 | 88.40 | **85.46** | 84.43 | 84.98 |
| Mixup-2017 [76] | 83.84 | 75.99 | 84.20 | 78.90 | 78.57 | 80.30 |
| BigAug-2020 [24] | 85.36 | 85.12 | 86.97 | 81.57 | 83.90 | 84.58 |
| Manifold Mixup-2019 [55] | 84.61 | 73.70 | 84.75 | 76.59 | 81.97 | 80.32 |
| Dofe-2020 [12] | 85.77 | 86.18 | 88.54 | 80.32 | 83.16 | 84.79 |
| Dofe w/ wrong domain label | 84.05 | 82.46 | 85.66 | 79.69 | 83.15 | 83.00 |
| CDDSA-2023 [13] | 85.22 | 82.27 | 87.10 | 81.99 | 80.90 | 83.50 |
| CDDSA w/ wrong domain label | 82.60 | 80.24 | 82.78 | 78.43 | 78.89 | 80.59 |
| TriD-2023 [61] | **89.16** | 82.78 | 86.62 | 84.81 | 83.54 | 86.35 |
| DFQ-2024 [62] | 88.54 | 85.04 | 88.63 | 85.31 | 85.10 | 86.82 |
| MixStyleFlow-2025 [64] | 85.42 | 83.19 | 86.04 | 83.19 | 82.18 | 84.00 |
| Aegis-2025 [65] | 86.39 | 86.58 | 87.91 | 84.63 | 85.84 | 86.27 |
| UniDDG w/o EMA | 86.68 | 86.87 | 87.48 | 80.29 | 86.37 | 85.54 |
| UniDDG w/o SA | 87.22 | 85.59 | 87.64 | 81.09 | **86.43** | 85.59 |
| **UniDDG (Ours)** | 87.65 | **88.01** | **89.16** | 83.70 | 86.30 | **86.96** |

As summarized in Table 8, UniDDG achieved the highest average Dice score of 86.96%, compared with 86.82% for DFQ and 86.27% for Aegis. Table 14 shows that the 0.14-percentage-point difference between UniDDG and the strongest comparison method, DFQ, was statistically significant ($p = 0.033$), demonstrating that the improvement was statistically detectable although modest in magnitude. For boundary accuracy, UniDDG achieved an average ASSD of 2.90 pixels, which was close to DFQ at 2.93 pixels and Aegis at 3.20 pixels, while Dofe obtained the lowest value of 2.87 pixels. These results show that UniDDG provides the strongest average Dice performance together with competitive boundary accuracy, rather than uniformly dominating every method on every metric.

Table 9 Evaluation of ASSD (pixel) ↓ using diverse DG approaches on prostate MRI dataset from multiple source centers to a single target center

| Method-Year | Center1 | Center2 | Center3 | Center4 | Center5 | Average |
|---|---|---|---|---|---|---|
| Lower bound w/o BDA | 3.95 | 6.58 | 6.21 | 6.26 | 2.65 | 5.13 |
| Lower bound | 3.29 | 3.86 | 4.92 | 3.89 | 2.48 | 3.69 |
| Upper bound w/o BDA | 2.62 | 4.03 | 3.50 | 2.94 | 4.25 | 3.47 |
| Upper bound | 3.56 | 3.95 | 2.18 | 7.43 | 3.04 | 4.03 |
| Cutmix-2019 [56] | 4.11 | 6.05 | 3.08 | **1.99** | 2.41 | 3.53 |
| Mixup-2017 [76] | 3.58 | 7.54 | 3.12 | 3.15 | 3.11 | 4.10 |
| BigAug-2020 [24] | 5.07 | 3.06 | 2.91 | 3.69 | 2.50 | 3.45 |
| Manifold Mixup-2019 [55] | 4.15 | 7.18 | 3.18 | 5.50 | 2.74 | 4.55 |
| Dofe-2020 [12] | 3.12 | 3.37 | **2.41** | 2.48 | 2.95 | **2.87** |
| Dofe w/ wrong domain label | 5.25 | 4.58 | 4.12 | 4.04 | 3.05 | 4.21 |
| CDDSA-2023 [13] | 4.57 | 5.29 | 4.43 | 2.12 | 4.78 | 4.24 |
| CDDSA w/ wrong domain label | 5.05 | 6.59 | 5.13 | 3.05 | 5.37 | 5.04 |
| TriD-2023 [61] | **2.12** | 4.62 | 3.18 | 2.97 | 2.82 | 3.14 |
| DFQ-2024 [62] | 2.79 | 3.99 | 3.20 | 2.48 | 2.21 | 2.93 |
| MixStyleFlow-2025 [64] | 5.12 | 4.90 | 3.00 | 2.26 | 2.89 | 3.63 |
| Aegis-2025 [65] | 2.86 | 3.01 | 3.14 | 4.58 | 2.42 | 3.20 |
| UniDDG w/o EMA | 2.77 | **2.92** | 2.91 | 3.87 | 2.20 | 2.93 |
| UniDDG w/o SA | 2.66 | 3.12 | 3.17 | 6.16 | 2.16 | 3.46 |
| UniDDG (Ours) | 2.63 | 4.20 | 2.65 | 2.88 | **2.14** | 2.90 |

Visual comparisons (Figure 8) further confirm that UniDDG achieves the largest overlap with the ground truth in the segmented regions, with fewer discrepancies in segmentation compared to other DG methods. These results on the prostate MRI dataset are consistent with the conclusions drawn from the fundus segmentation experiments, further highlighting the generalizability and reliability of UniDDG. By achieving superior performance in both segmentation accuracy and boundary precision, UniDDG demonstrates its effectiveness as a robust framework for generalizing from multiple source centers to a single target center in medical image segmentation tasks, making it highly applicable to real-world clinical scenarios.

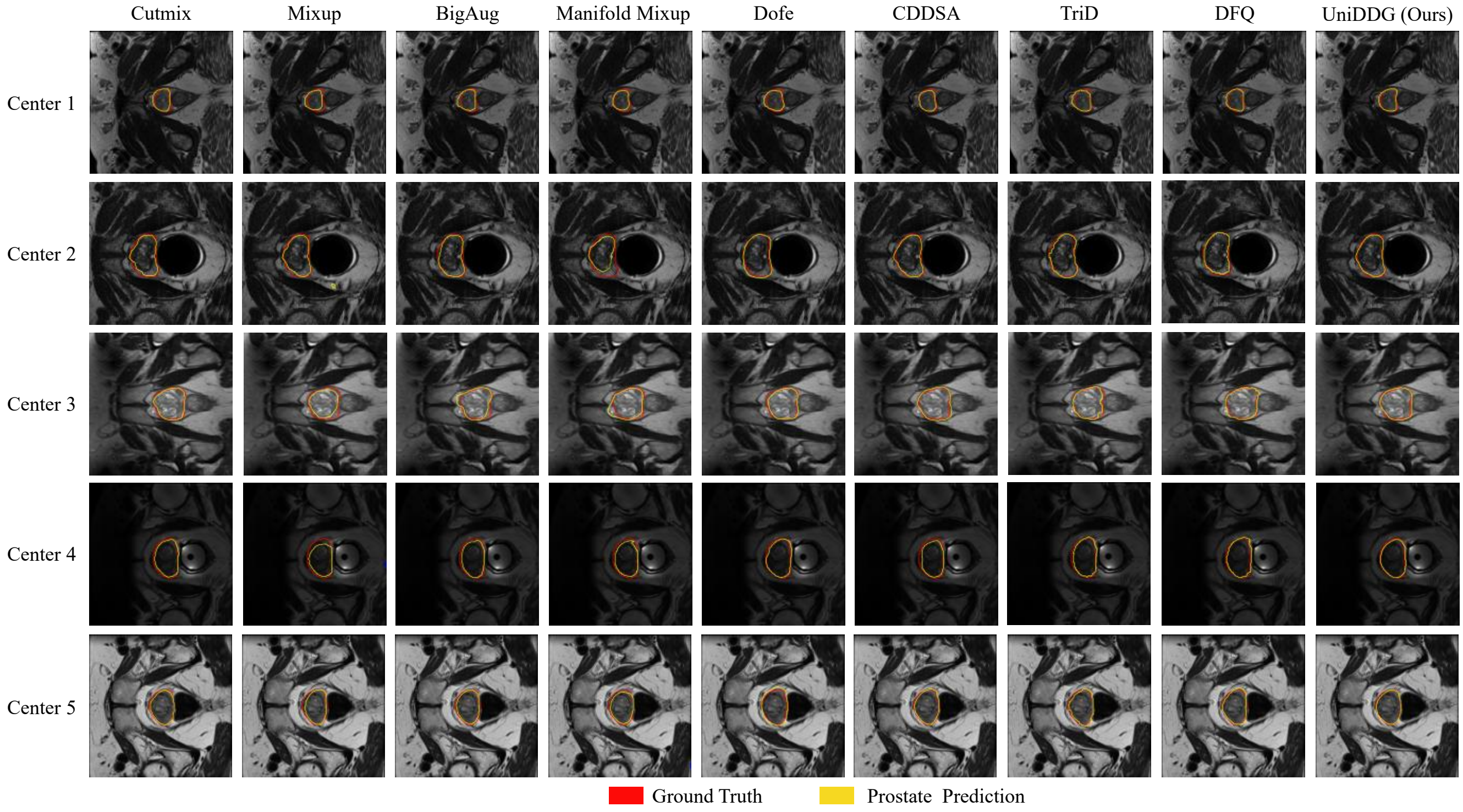

Figure 8 Visualization of segmentation results using diverse DG approaches on prostate MRI dataset from multiple source centers to a single target center

(ii) Ablation studies: We conducted ablation studies by removing the EMA and SA modules, the results summarized in Table 8 and Table 9. When the EMA module was removed (UniDDG w/o EMA), there was a noticeable drop in performance. Specifically, the average Dice score decreased from 86.96% to 85.54%, a reduction of 1.42%, and the average ASSD increased from 2.90 pixels to 2.93 pixels. Similarly, removing the SA module (UniDDG w/o SA) also led to a performance decline, although the impact was less pronounced compared to the removal of the EMA module. The average Dice score decreased from 86.96% to 85.59%, and the average ASSD increased from 2.90 pixels to 3.46 pixels. The results from these ablation studies indicate that both EMA and SA significantly contribute to the overall performance of UniDDG.

### *4.4.2. Generalization within a single center*

(i) Comparison with state-of-the-art DG methods: To investigate the performance of UniDDG in the context of a single target center, we conducted experiments using a prostate MRI dataset, following the same evaluation protocol as in the section 4.3.2. In this case, the generalization task involves training on data from a single center and testing on data from the same center, which is more representative of real-world clinical scenarios where data from a specific medical facility is used for model development.

Table 10 Evaluation of Dice Scores (%) ↑ using diverse DG approaches on prostate MRI dataset within a single center

| Method-Year | Center1 | Center2 | Center3 | Center4 | Center5 | Average |
|---|---|---|---|---|---|---|
| Lower bound w/o BDA | 88.08 | 86.02 | 86.59 | 83.00 | 77.47 | 84.23 |
| Lower bound | 88.00 | 88.53 | 89.61 | 81.74 | 78.45 | 85.27 |
| Cutmix-2019 [56] | 88.59 | 88.55 | 90.66 | 86.56 | 76.27 | 86.13 |
| Mixup-2017 [76] | 87.64 | 83.97 | 88.43 | 81.88 | 76.32 | 83.65 |
| BigAug-2020 [24] | 88.47 | 86.02 | 89.44 | 83.16 | 75.51 | 84.52 |
| Manifold Mixup-2019 [55] | 88.10 | 85.22 | 88.49 | 82.21 | 73.21 | 83.45 |
| DAPSAM-2024 [63] | 88.68 | 89.47 | 89.85 | 86.62 | 80.90 | 87.10 |
| UniDDG w/o EMA | 88.38 | 89.57 | 90.65 | 87.38 | 78.52 | 86.90 |
| UniDDG w/o SA | 88.66 | 90.21 | 90.72 | 88.17 | 81.07 | 87.77 |
| **UniDDG (Ours)** | **89.09** | **91.36** | **90.93** | **89.23** | **82.20** | **88.56** |

As shown in Table 10 and Table 11, UniDDG consistently outperforms the other domain generalization (DG) methods in both Dice score and ASSD. Specifically, UniDDG achieves the highest average Dice score of 88.56%, surpassing the best-performing baseline method, Cutmix, by 2.43%. It is also worth noting that UniDDG outperforms DAPSAM (87.10%), which utilizes domain-adaptive prototypes. This indicates that for specific medical segmentation tasks where data volume is limited (as in single-center prostate MRI), our efficient disentanglement framework can achieve superior results compared to prompt-learning approaches that may require larger-scale pre-training or adaptation. Additionally, in terms of boundary precision, UniDDG achieves the lowest average ASSD of 2.56 pixels, outperforming Cutmix (3.08 pixels) and all other methods. This indicates that UniDDG excels in both overall segmentation accuracy and fine boundary delineation, making it a highly effective framework for generalizing medical image segmentation tasks within a single center. The results further demonstrate UniDDG's ability to maintain segmentation performance across different centers, showcasing its robustness in handling variations in medical imaging data.

Table 11 Evaluation of ASSD (pixel) ↓ using diverse DG approaches on prostate MRI dataset within a single center

| Method-Year | Center1 | Center2 | Center3 | Center4 | Center5 | Average |
|---|---|---|---|---|---|---|
| Lower bound w/o BDA | 2.62 | 4.03 | 3.50 | 2.94 | 4.25 | 3.47 |
| Lower bound | 3.56 | 3.95 | 2.18 | 7.43 | 3.04 | 4.03 |
| Cutmix-2019 [56] | **2.54** | 2.75 | 2.12 | 2.24 | 5.77 | 3.08 |
| Mixup-2017 [76] | 2.73 | 4.42 | 2.23 | 3.26 | 6.16 | 3.76 |
| BigAug-2020 [24] | 3.43 | 4.03 | 2.13 | 3.02 | 8.30 | 4.18 |
| Manifold Mixup-2019 [55] | 3.35 | 4.05 | 2.40 | 3.10 | 6.52 | 3.88 |
| DAPSAM-2024 [63] | 3.10 | 2.82 | 2.87 | **1.85** | **2.65** | 2.66 |
| UniDDG w/o EMA | 3.48 | 2.56 | 2.13 | 1.94 | 3.95 | 2.81 |
| UniDDG w/o SA | 3.46 | 2.45 | 2.00 | 4.35 | 3.03 | 3.06 |
| **UniDDG (Ours)** | 3.21 | **2.16** | **1.92** | 2.13 | 3.39 | **2.56** |

ii) Ablation studies: We conducted ablation studies by removing the EMA and SA modules, The results of these experiments are summarized in Table 10 and Table 11. When the EMA module is removed (UniDDG w/o EMA), the performance deteriorates significantly. Specifically, the average Dice score decreased from 88.56% to 86.90%, and the average ASSD increased from 2.56 to 2.81 pixels. Similarly, when the SA module is removed (UniDDG does not carry SA), there is also a decrease in performance, although to a lesser extent. The average Dice score decreased from 88.56% to 87.77%, and the average ASSD increased from 2.56 to 3.06 pixels. The results of these ablation studies indicate that both EMA and SA contribute significantly to the overall performance of UniDDG.

A comparison across the four experimental settings in Tables 4–11 further clarifies the relative contributions of the two independently ablated components. Removing EMA reduces the average Dice score by 2.33, 0.99, 1.42, and 1.66 percentage points in the fundus multi-source, fundus single-source, prostate multi-source, and prostate single-source settings, respectively. Removing SA reduces Dice by 0.88, 0.32, 1.37, and 0.79 percentage points in the same settings. EMA therefore provides the larger incremental Dice contribution in all four settings. SA also improves Dice consistently, although its effect on ASSD depends on the evaluation setting: removing SA slightly improves ASSD in the fundus multi-source experiment, whereas the full model achieves better ASSD in the other three settings. The OIOD-guided representation exchange and consistency constraints described in Section 3.1 remain common to all three variants and constitute the shared objective-level mechanism. Accordingly, the existing ablations identify EMA as the stronger incremental contributor to Dice and SA as a smaller but consistent contributor.

The experimental evidence is organized at two levels. Tables 4–11 compare the complete UniDDG framework with standard augmentation and DG methods, while the multi-source experiments in Tables 4–5 and 8–9 additionally compare it with center-label-dependent methods, including Dofe and CDDSA. These are framework-level comparisons between alternative DG paradigms. Within UniDDG, the w/o EMA and w/o SA variants provide controlled pathway-level ablations: removing EMA eliminates the masked reconstruction pathway, whereas removing SA eliminates the random-style branch and its content- and prediction-consistency constraints. By contrast, the OIOD formulation, content-style disentanglement, cross-image style exchange, and their consistency constraints constitute the coupled core shared by all UniDDG variants. Replacing image-wise style exchange with center-defined grouping would define a different training objective rather than remove an independent add-on. Therefore, Tables 4–11 provide framework-level evidence for the complete OIOD-guided framework and controlled component-level evidence for its EMA and SA pathways.

To further substantiate the robustness of the EMA module, we conducted a sensitivity analysis regarding the expansion distance used for mask generation. We evaluated the segmentation performance on Center 1 of the prostate dataset by systematically varying the expansion radius from 3 to 20 pixels. The quantitative results are presented in Table 12.

Table 12 Sensitivity analysis of Expansion Mask Attention (EMA) distance on Center 1 (Prostate).

| Expansion Distance (Radius) | Dice (%) | ASSD |
|---|---|---|
| 3 pixels | 86.92 | 2.85 |
| 5 pixels | 87.35 | 2.68 |
| **10 pixels** | **87.65** | **2.63** |
| 15 pixels | 87.48 | 2.71 |
| 20 pixels | 86.83 | 2.79 |

As illustrated in Table 12, the model maintains high stability within a radius range of 5 to 15 pixels. Specifically, the Dice score peaked at 87.65% with a radius of 10 pixels. Performance exhibited a minor decline when the radius was too small (3 pixels) due to insufficient contextual information, or too large (20 pixels) due to the inclusion of excessive background noise. This confirms that while the expansion distance plays a role in boundary attention, the UniDDG framework is not hypersensitive to this specific hyperparameter.

To further substantiate the superiority of the proposed EMA strategy over traditional explicit edge supervision methods, we compared UniDDG with variants using Sobel Edge Loss and Canny Edge Loss on the Center 1 dataset. We hypothesized that explicit edge detectors, which rely on pixel-level gradients, would be less robust to the noise and contrast variations inherent in MRI images compared to the region-based attention of EMA. The comparative results are presented in Table 13.

As shown in Table 13, while explicit edge losses provide some improvement over the baseline by enforcing high-frequency consistency, they consistently underperform compared to EMA. This confirms that gradient-based methods (Sobel) and binary edge methods (Canny) are sensitive to local intensity shifts and noise, potentially enforcing incorrect "noise edges." In contrast, EMA's regional attention provides a softer, semantically driven constraint that is more robust to domain-specific contrast variations.

Table 13 Comparison of EMA strategy against traditional explicit edge losses on Center 1 (Prostate).

| Method | Mechanism | Dice (%) | ASSD |
|---|---|---|---|
| w/o EMA | No boundary attention | 86.68 | 2.77 |
| w/ Sobel Loss | Gradient alignment | 86.93 | 2.81 |
| w/ Canny Loss | Binary edge alignment | 87.10 | 2.68 |
| **w/ EMA (Ours)** | **Regional mask attention** | **87.65** | **2.63** |

### 4.5. Statistical significance analysis

We further compared UniDDG with the strongest comparison method in each of the four experimental settings. Table 14 reports the average Dice scores, absolute differences, and corresponding $p$ values.

Table 14 Statistical comparison between UniDDG and the strongest comparison method in each experimental setting.

| Dataset | Setting | Comparison method | UniDDG Dice (%) | Comparison Dice (%) | Difference (percentage points) | p value |
|---|---|---|---|---|---|---|
| Fundus | Multi-source | DFQ | 84.43 | 84.36 | +0.07 | 0.043 |
| Fundus | Single-center | DAPSAM | 88.91 | 88.80 | +0.11 | 0.034 |
| Prostate | Multi-source | DFQ | 86.96 | 86.82 | +0.14 | 0.033 |
| Prostate | Single-center | DAPSAM | 88.56 | 87.10 | +1.46 | <0.001 |

Positive differences indicate higher Dice scores for UniDDG. All four comparisons reached statistical significance at $\alpha = 0.05$. The absolute differences in the first three settings were small, ranging from 0.07 to 0.14 percentage points, whereas the single-center prostate setting showed a larger difference of 1.46 percentage points. These results indicate that the Dice advantage of UniDDG was statistically detectable in all four evaluated settings, while the magnitude of the improvement depended on the dataset and evaluation protocol. The sample size of each test is stated in Section 4.2 (N = 271 for the fundus settings and N = 22 for the prostate settings).

### 4.6. Multi-source domain generalization under misassigned explicit domain labels

In DG tasks, methods such as Dofe and CDDSA often rely heavily on explicit domain labels to define the source domain of model adaptation during training. These methods assume that domain labels are correctly assigned and used during the training phase. In practice, however, domain labels can be misassigned or even unavailable, which can seriously harm model performance. In contrast, our proposed UniDDG model does not rely on explicit domain labels based on the "one image is one domain" (OIOD) assumption. It treats each image as a separate domain, mixing all source domain data together during training without the need for explicit label assignment. This randomization reduces the dependence on domain labels and enhances the robustness of the model to domain shifts.

To evaluate the degree of dependency on explicit domain labels, we conducted an experiment where we shuffled domain labels across multiple source centers. Specifically, we mixed the data from different centers into a single image pool and then randomly reassigned domain labels (as shown in Figure 9). After shuffling the labels, each "new domain" contained the same amount of data as the original, but with incorrect label assignments. We then used t-SNE to visualize the feature distributions of the training data before and after label shuffling (Figure 10).

Figure 10 visualizes the same VGG16 feature embeddings under two different label assignments; therefore, shuffling the labels does not alter the underlying feature geometry. Under the original center labels, the labels retain a partial correspondence with the observed feature neighborhoods. After label shuffling, samples assigned to the same artificial domain become dispersed across the feature space, while nearby samples may receive different domain labels. Together with the substantial intra-center variation and inter-center overlap shown in Figure 1, this result demonstrates that externally assigned domain labels may impose grouping relationships that do not accurately represent image-level appearance similarities. The performance degradation of Dofe and CDDSA reported in Table 4, Table 5, Table 8, and Table 9 further shows that encoding such inaccurate relationships into the training objective can impair generalization. It should be noted that Figure 10 is an illustrative visualization of the potential mismatch between externally assigned labels and the underlying feature geometry; the causal evidence is provided by the controlled label-shuffling comparison in this section, in which the data and architecture are identical and only the labels differ.

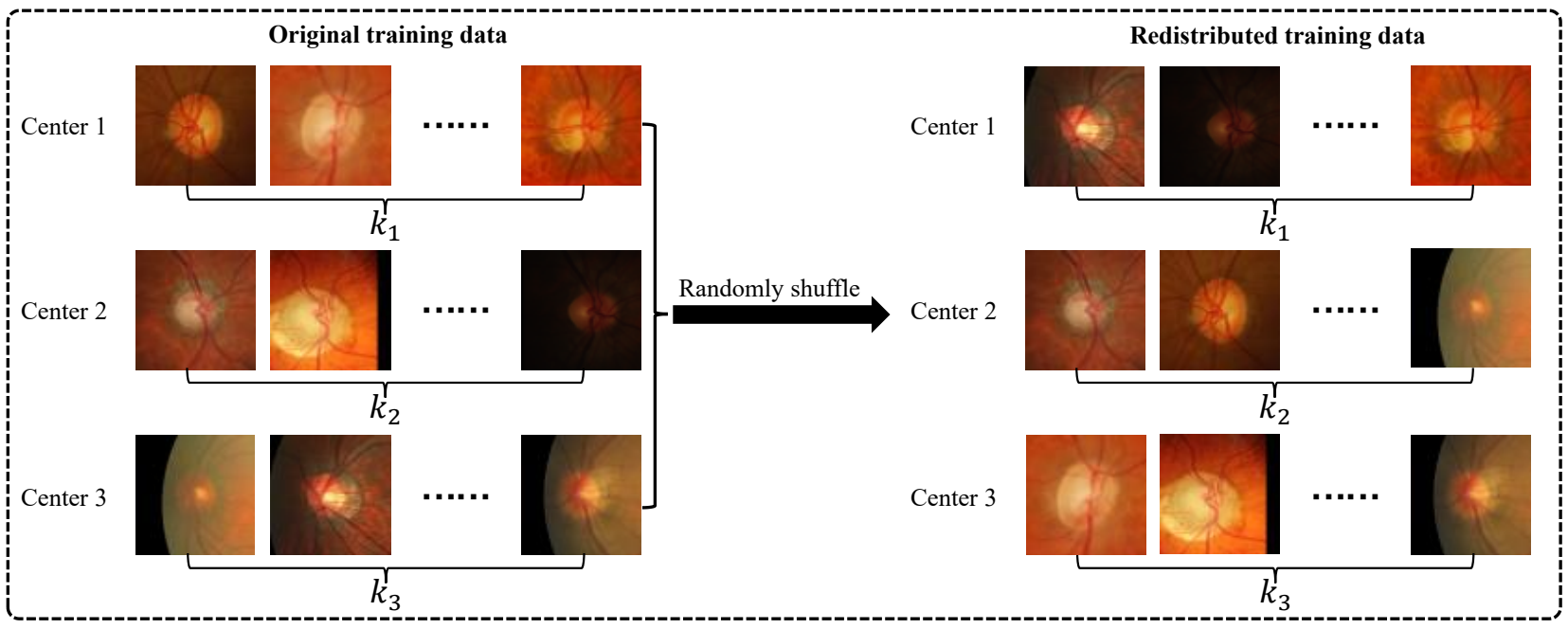

Figure 9 Illustration of shuffling domain labels of multi-source central data

We then trained Dofe and CDDSA on the shuffled data using the incorrect domain labels and evaluated their generalization performance on a new target center. The results are presented in Table 4, Table 5, Table 8, and Table 9. Table 4 and Table 5 involve the segmentation of the optic cup and optic disc from fundus images, while Table 8 and Table 9 focus on prostate MRI image segmentation tasks. The results for Dofe and CDDSA with shuffled domain labels are labeled as Dofe w/ wrong domain label and CDDSA w/ wrong domain label. As shown in these tables, both methods exhibited significant performance degradation when domain labels were incorrect.

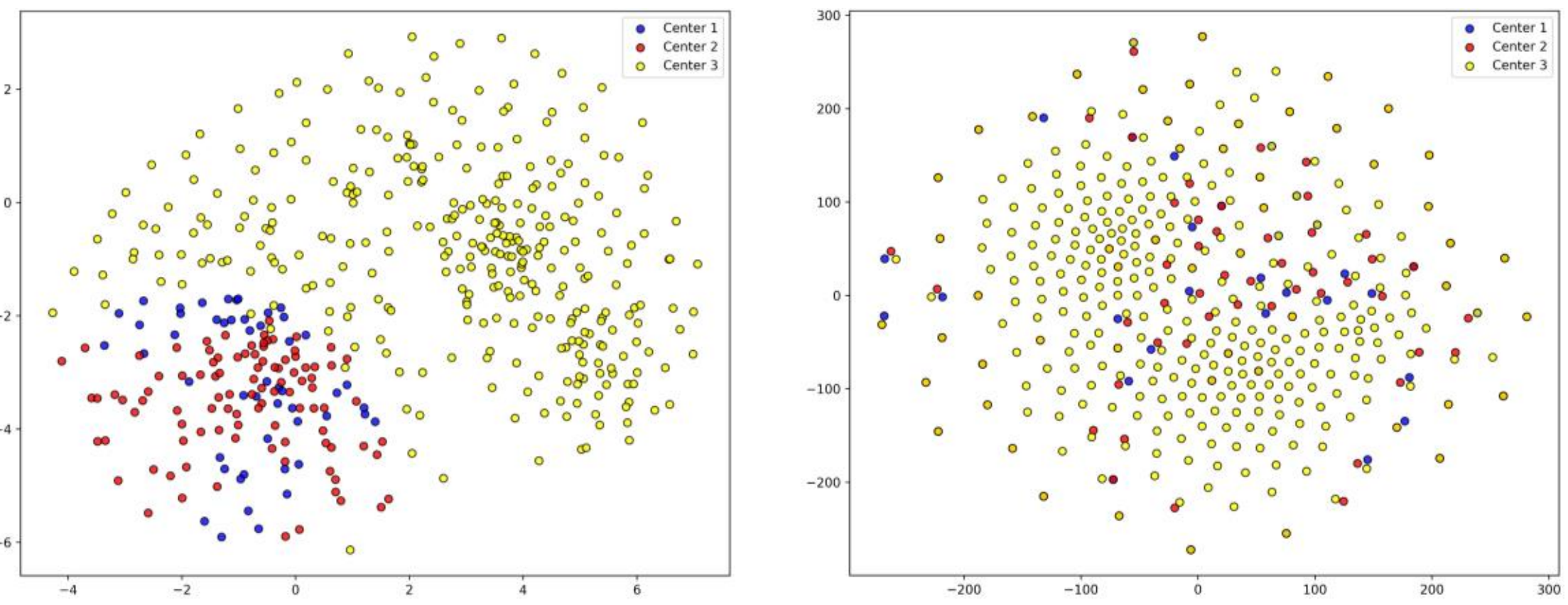

(a) Feature distribution of original training datasets (b) Feature distribution of shuffled training datasets

Figure 10 t-SNE visualization of VGG16 features of fundus images from original and shuffled training datasets for OC/OD segmentation

For example, in Table 4 (Dice scores for fundus image segmentation), Dofe's performance dropped from 83.91% to 83.26% with shuffled labels, while CDDSA showed a more substantial decrease from 82.84% to 80.80%. In Table 5 (ASSD for fundus image segmentation), Dofe's ASSD increased from 3.45 to 3.85, while CDDSA's ASSD rose sharply from 3.65 to 4.25. Similarly, in Table 8 and Table 9, where the segmentation task involves prostate MRI images, the performance of Dofe and CDDSA significantly worsened when domain labels were shuffled. This confirms that both Dofe and CDDSA rely heavily on explicit domain labels during training. The performance drop was especially pronounced in CDDSA, suggesting that this method is more sensitive to misassigned domain labels compared to Dofe.

The stronger dependence of CDDSA on domain labels can be attributed to the "Domain Style Contrastive Learning" technique included in this approach. This technique requires domain labels to effectively separate the features of different domains. This method forces the model to separate features from different domains while pulling features from the same domain closer together in the feature space. This contrast learning approach relies heavily on explicit domain labels, making CDDSA more sensitive to label misassignment than Dofe.

In contrast, our UniDDG method does not rely on explicit domain labels and is therefore relatively immune to the misassignment of domain labels. Since UniDDG's training process intentionally eliminates the dependence on domain labels, the performance degradation observed in Dofe and CDDSA when domain labels are shuffled is avoided. This highlights the robustness of UniDDG's domain-independent approach, which performs well even when domain labels are incorrectly assigned, making it more suitable for real-world scenarios where domain labels may be unreliable or unavailable.

Direct validation of the OIOD formulation. To directly attribute the gains of UniDDG to the image-level domain formulation, we implemented two controlled variants of the same framework with all other components (content encoder, style encoder, decoder, segmenter, EMA, SA, data, and evaluation protocol) held fixed: (1) a center-grouped variant, in which style codes are grouped and exchanged according to source-center membership and a contrastive term imposes within-center compactness and between-center separation on the style space, as in CDDSA[13]; and (2) an unconstrained-exchange variant, in which style codes are exchanged globally and random style augmentation is applied without the content, style, or prediction consistency constraints. Table 15 reports the average Dice on the multi-source fundus and prostate tasks. The center-grouped and unconstrained variants attain 83.61% and 81.73% on fundus, and 86.02% and 85.46% on prostate, respectively, compared with 84.43% and 86.96% for the proposed OIOD formulation. Variants (1) and (2) differ in both the partition criterion and the consistency constraints, and the center-grouped arm additionally carries a contrastive term; their differences therefore cannot be attributed to a single factor. We interpret the three-arm pattern as converging evidence rather than as a factorial decomposition: both alternative formulations, center-grouped (83.61%/86.02%) and unconstrained exchange (81.73%/85.46%), fall below the proposed formulation (84.43%/86.96%), which shows that neither center-level partitioning nor unconstrained exchange reproduces the benefit of the image-level formulation; together they provide converging evidence that the gains follow from the image-level formulation.

Effect of the KL divergence. To directly quantify the effect of removing the KL regularization toward the unit Gaussian prior (Section 3.3), we re-introduced it with weight 1.0 while keeping all other coefficients fixed. With KL regularization, the average Dice is 84.12% on fundus and 85.87% on prostate, compared with 84.43% and 86.96% for the reported configuration, confirming that the image-conditioned representation used by UniDDG performs at least as well as pushing all style codes toward a shared Gaussian prior.

Table 15 Controlled comparison of the domain formulation within the same UniDDG objective, together with the effect of re-introducing the KL divergence.

| Settings | Fundus (avg. Dice %) | Prostate (avg. Dice %) |
|---|---|---|
| (1) Center-grouped formulation | 83.61 | 86.02 |
| (2) Unconstrained random exchange | 81.73 | 85.46 |
| **Proposed OIOD (ours)** | **84.43** | **86.96** |
| OIOD + KL (weight = 1.0) | 84.12 | 85.87 |

Sensitivity of the main loss coefficients. To examine robustness to the loss weighting, we perturbed the three coefficients with the largest role — λ1 (segmentation), λ3 (masked reconstruction), and λ5 (style consistency) — by ±50%, on the multi-source fundus and prostate settings, with all other coefficients fixed at their reported values. Across these perturbations, the average Dice varies within 1 percentage points, indicating that the reported configuration is not an isolated peak in the loss-weighting landscape.

Table 16 Sensitivity analysis of the main loss coefficients.

| Settings | Fundus (avg. Dice %) | Prostate (avg. Dice %) |
|---|---|---|
| **Ours (λ1=5.0, λ3=10.0, λ5=5.0)** | **84.43** | **86.96** |
| λ1× 0.5 / × 1.5 | 84.17 / 84.02 | 86.74 / 86.62 |
| λ3 × 0.5 / × 1.5 | 84.04 / 84.15 | 86.53 / 86.69 |
| λ5 × 0.5 / × 1.5 | 84.12 / 84.21 | 86.79 / 86.71 |

### 4.7. Error Analysis

To provide a balanced evaluation of UniDDG, we analyzed failure cases where the model struggled to produce accurate segmentations. Figure 11 illustrates two such examples from the test set. In the first case (Optic Disc/Cup segmentation), the image exhibits extremely low contrast with severe blurring around the optic cup boundary. The model fails to distinguish the cup from the disc, leading to under-segmentation. In the second case (Prostate segmentation), the MRI slice contains significant motion artifacts and noise, which disrupts the boundary continuity of the prostate gland. Consequently, the model produces a fragmented mask. These cases suggest that while UniDDG is robust to style shifts, its performance can still be compromised by severe image quality degradation or ambiguous anatomical features, which remains a challenge for future improvement.

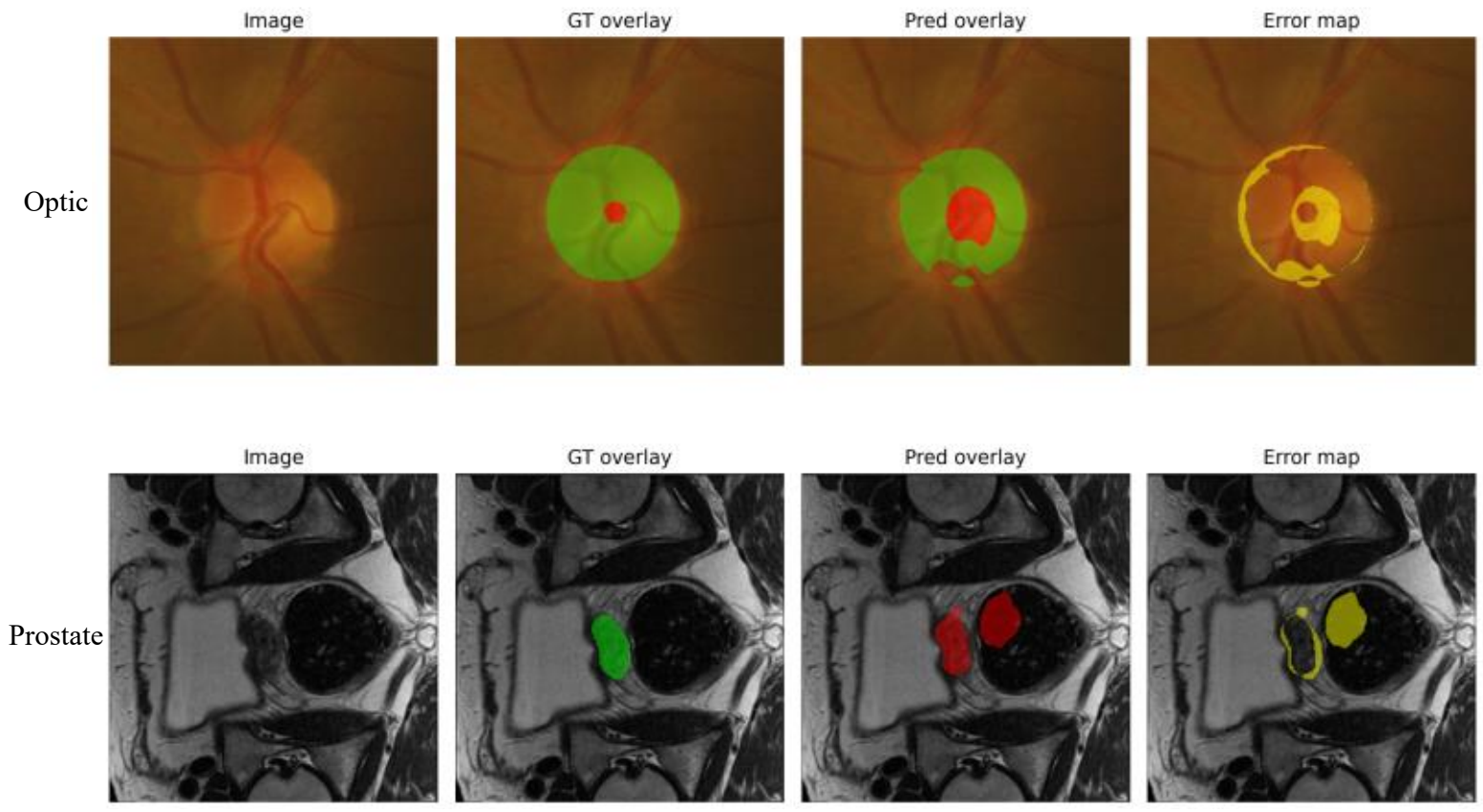

Figure 11 Visualization of failure cases. Top row: An optic disc/cup sample with low contrast causing boundary error. Bottom row: A prostate sample with artifacts causing segmentation fragmentation

To further examine whether the disentanglement objective behaves as intended, Figure 12 shows t-SNE projections of the learned style codes and of the pooled content features of the held-out test images, colored by source-center membership. The style codes are, to a certain extent, separated by center, indicating that they retain the center-related acquisition appearance. In contrast, the content features overlap substantially across centers, with localized residual cluster structure visible in the two-dimensional projection, providing direct evidence that the anatomical content has been effectively disentangled from acquisition-related style and is therefore domain-invariant.

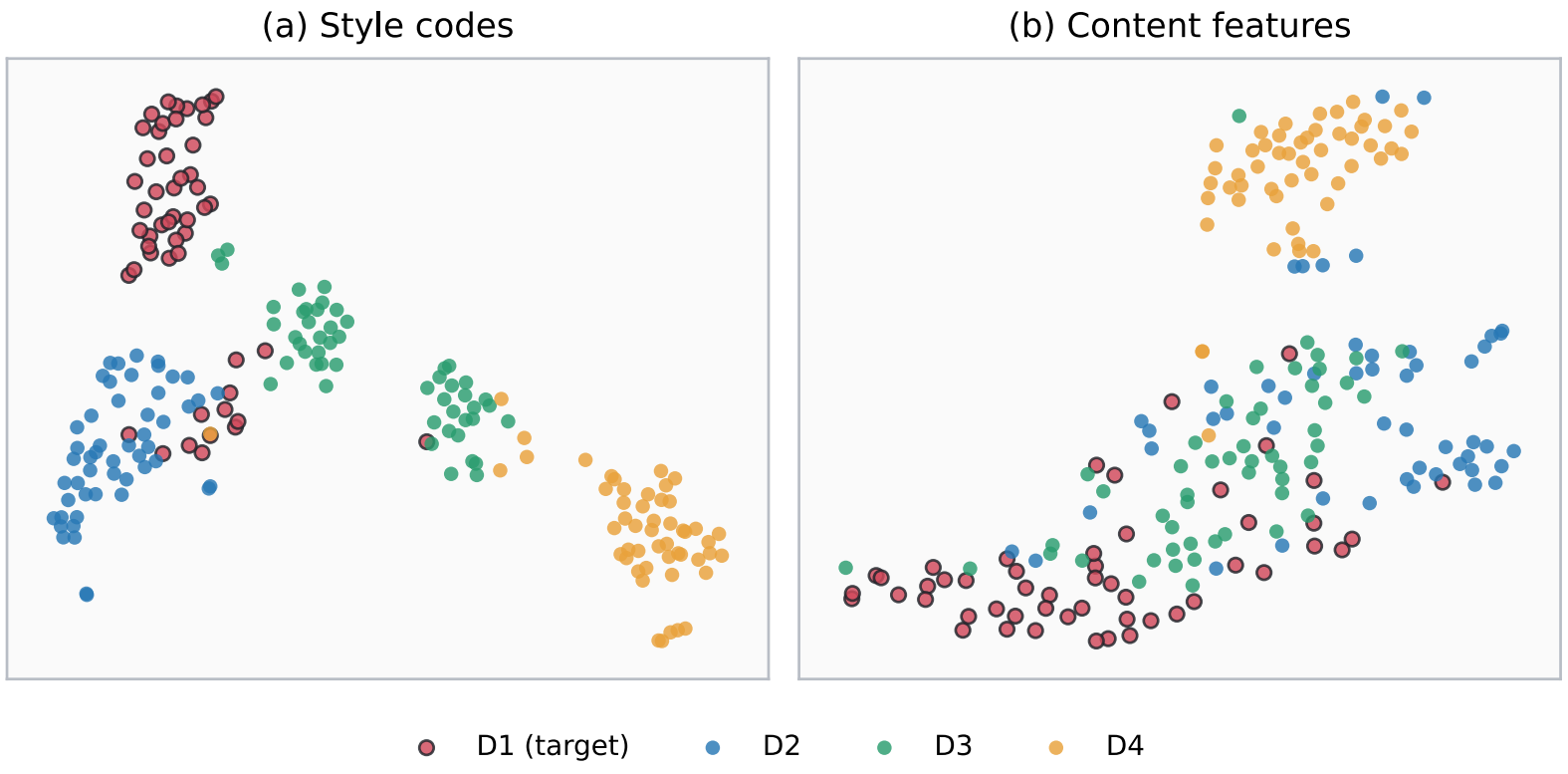

Figure 12 t-SNE visualization of the learned style codes and content features of held-out test images

## 5. Discussion

### 5.1. Interpretation of the main findings

The experimental results reveal three main patterns. First, UniDDG applies the same image-level domain formulation to both multi-source and single-center training without requiring center labels or setting-specific architectural changes. Table 14 shows statistically significant Dice improvements in all four evaluated settings, although the effect sizes differ substantially. The improvements in the first three settings range from 0.07 to 0.14 percentage points, whereas the single-center prostate setting shows a larger improvement of 1.46 percentage points. Thus, the results support consistent applicability across the evaluated protocols rather than uniformly large performance margins.

The results provide empirical support for removing center identity from the disentanglement objective. Figure 1 shows substantial style variation within individual centers and overlap between centers, indicating that center membership is an imperfect proxy for image style. Under OIOD, image-specific style codes are globally reassigned while the content, style, and prediction consistency losses require anatomical information to remain unchanged. This encourages the content representation to remain stable across a broader set of style variations than those defined by center-level grouping. The domain-label reassignment experiment in Section 4.6 further supports the practical value of this design: Dofe and CDDSA deteriorate when their center labels are incorrect, whereas UniDDG does not consume these labels during training.

Computational scalability and source-center annotation requirements distinguish the evaluated frameworks along separate axes. CDDSA [13] also uses a shared architecture whose parameter count does not increase with the number of source centers. A fixed model size is therefore not unique to UniDDG. UniDDG combines this architectural property with image-level style modeling that excludes center identity from style supervision, pair construction, and style exchange. Table 3 reports 3.44 M parameters and 26.76 GFLOPs per image during training. Removing the style encoder and decoder reduces the inference model to 2.20 M parameters and 8.54 GFLOPs. The training and inference computation graphs remain $O(1)$with respect to the number of source centers ($K$). Section 4.6 shows that Dofe and CDDSA lose accuracy when center labels are misassigned, whereas UniDDG does not use these labels. Figure 1 provides the data-level motivation by showing substantial variation within individual centers and overlap between centers. This comparison positions OIOD as the combination of image-level variability modeling, independence from source-center annotations, and fixed architectural complexity.

We do not claim a formal proof that the image-level partition generalizes better than center-level groupings; the comparison mechanisms in this field — style randomization, contrastive disentanglement, adversarial training, and normalization-based invariance — are likewise validated empirically rather than formally. The argument for OIOD is twofold. Empirically, Figure 1 shows that intra-center variation is comparable to inter-center variation and that center clusters overlap, so center membership is an unreliable partition of style. Conceptually, if anatomical content is invariant to acquisition style, then training the content encoder over a broader and denser observed style support — the image-level support induced by OIOD, which spans both intra- and inter-center variation — reduces the expected divergence to unseen acquisition styles relative to a coarser partition that is partly inconsistent with observed appearance. We therefore present OIOD as a conceptually motivated and empirically validated formulation rather than as a formally proven bound.

Recent foundation models, such as the Segment Anything family, provide powerful general-purpose segmenters, and several recent works adapt them to unseen domains through prompting or prototype conditioning [63]. UniDDG is complementary to this line: it trains a lightweight disentanglement-based segmenter from scratch on the available source data, does not depend on large pretrained backbones or additional prompt tuning, and targets the domain formulation rather than the segmentation backbone. A quantitative comparison with foundation-model-based methods is deferred until standardized evaluation protocols for generalizable medical image segmentation become available (Section 6).

### 5.2. Component behavior and failure modes

The ablation results indicate that EMA and SA address different aspects of generalization. EMA constrains reconstruction around anatomically relevant regions and prevents the decoder from introducing artificial boundary cues that can become shortcuts for the segmentation network. SA expands the styles encountered during training and improves adaptation to acquisition-related appearance changes. Their effects are nevertheless metric-dependent. For example, UniDDG achieves the highest average Dice score on the multi-source fundus experiment, but its ASSD remains higher than that of DFQ and Aegis. This suggests that preserving regional anatomical consistency does not necessarily optimize contour distance to the same extent as methods that directly emphasize boundary localization.

The failure cases in Figure 11 further clarify the operating boundary of the method. Severe blur can remove the contrast needed to distinguish the optic cup from the disc, while motion artifacts can interrupt the visible continuity of the prostate boundary. In these cases, the missing or corrupted anatomical evidence cannot be recovered through style exchange alone. UniDDG is therefore most effective when domain shifts primarily alter acquisition-dependent appearance while the underlying anatomical structure remains observable. This interpretation explains both the gains under center-related style variation and the remaining errors under severe image-quality degradation.

## 6. Limitations

First, the empirical scope of this study comprises two segmentation tasks, two imaging modalities, and nine center-specific datasets. These data capture substantial variations in acquisition sites, scanners, magnetic field strengths, resolutions, and imaging protocols, but they do not represent the full diversity of medical imaging. In particular, the current evidence does not establish performance on CT, ultrasound, histopathology, multi-organ segmentation, or diffuse and multifocal targets. Evaluation on additional modalities and anatomical structures is therefore necessary to determine how broadly the observed center-related generalization transfers.

OIOD assumes that image-specific appearance can be treated as a style factor that is separable from anatomical content. This assumption is suitable when domain shifts primarily arise from acquisition-dependent contrast, illumination, noise, or scanner characteristics. However, style and anatomy may become correlated when acquisition protocols are associated with particular patient populations, diseases, or anatomical conditions. Global style exchange could then generate combinations that are uncommon in the original data or treat pathological variation as a nuisance factor. The reconstruction and consistency objectives constrain such exchanges to preserve anatomy, but they do not constitute a formal guarantee of complete content-style disentanglement.

All experiments use a two-dimensional U-Net-based content encoder. The prostate MRI slices are processed independently and their grayscale channel is replicated to match the three-channel input used for the fundus task. Although this shared implementation enables controlled comparisons across the two modalities, it does not exploit three-dimensional context and has not been validated with nnU-Net variants, transformer-based encoders, or other modern segmentation backbones. Consequently, the current results establish the effectiveness of UniDDG with the reported backbone rather than backbone-independent performance.

The evaluation focuses on Dice and ASSD and does not assess probability calibration, predictive uncertainty, or prospective clinical deployment. The computational analysis in Table 3 reports architecture-derived parameters, FLOPs, and parameter storage, but not wall-clock training time or peak activation memory. Future evaluation should therefore include calibrated uncertainty estimates, prospective or external validation, and standardized runtime profiling before the framework is considered for deployment-oriented use. In addition, quantitative comparison with recent foundation-model-based segmentation methods should be performed once standardized evaluation protocols for generalizable medical image segmentation become available.

## 7. Conclusion

In this paper, we developed a new perspective on domain generalization for medical image segmentation, which challenges the conventional practice of grouping images from the same center or scanner as a single domain. A new hypothesis named "One image as a Domain" (OIOD) was presented, which means treating each image as a separate domain, acknowledging inherent variability within and across centers. This enables more flexible and robust generalization because it takes into account inter-center and intra-center domain shifts which are often overlooked in traditional domain generalization methods.

Based on the OIOD hypothesis, we propose the UniDDG framework, which is a unified disentanglement-based domain generalization method that can handle both single and multi-source domain generalization without the need for explicit domain labeling. UniDDG decouples content representation and style code through operations such as exchange, recombination, and reconstruction. At inference, it adds no computational overhead relative to the backbone, because the style encoder and decoder are discarded (2.20 M parameters and 8.54 GFLOPs per image); during training, these auxiliary modules introduce additional parameters and FLOPs (1.24 M parameters and 18.22 GFLOPs per image), as quantified in Table 3, and the parameter count of the deployed network remains the same regardless of the number of source domains. In addition, UniDDG combines Expansion Mask Attention (EMA) to preserve boundaries during reconstruction and Style Augmentation (SA) to simulate different image styles, thus enhancing the robustness of the model to domain shift. Extensive experiments on multi-source and single-source medical image segmentation tasks were conducted, including optic disc and cup segmentation as well as prostate segmentation. Those experiments demonstrate that our method outperforms several existing state-of-the-art methods.

### Data and code availability statement

Data and code are made available at https://github.com/Captain-Hong/UniDDG.

### Declaration of competing interest

The authors declare that there are no conflicts of interest regarding the publication of this paper.

### Acknowledgements

This work was supported in part by the National Natural Science Foundation of China (62466033), in part by the Jiangxi Provincial Natural Science Foundation (20242BAB20070).